\documentclass{article}

\usepackage[a4paper,margin=25mm]{geometry}
\usepackage{microtype}
\usepackage{graphicx}
\usepackage{amsmath,amssymb,amsfonts}
\usepackage{bm}
\usepackage{physics}
\usepackage{siunitx}
\usepackage{authblk}
\usepackage[numbers,sort&compress]{natbib}
\usepackage[colorlinks=true,linkcolor=blue,citecolor=blue,urlcolor=blue]{hyperref}
\usepackage{caption}
\usepackage{subcaption}
\usepackage{booktabs}
\usepackage{placeins}
\usepackage{xcolor}
\usepackage{indentfirst}
\usepackage{float}

\title{Design of a radio-frequency photoinjector for generating relativistic vortex electron beams}
\author{A.S.~Dyatlov\textsuperscript{1,2,\thanks{email:~\href{mailto:aleksandr.dyatlov@metalab.ifmo.ru}{aleksandr.dyatlov@metalab.ifmo.ru}}},
V.V.~Kobets\textsuperscript{2},
A.E.~Levichev\textsuperscript{3},
M.V.~Maksimov\textsuperscript{1},
D.A.~Nikiforov\textsuperscript{3},
M.A.~Nozdrin\textsuperscript{2,\thanks{email:~\href{mailto:nozdrin@jinr.ru}{nozdrin@jinr.ru}}},
K.~Popov\textsuperscript{2,4,\thanks{email:~\href{mailto:popovkon@jinr.ru}{popovkon@jinr.ru}}},
K.A.~Sibiryakova\textsuperscript{3},
K.E.~Yunenko\textsuperscript{2},
D.V.~Karlovets\textsuperscript{1,5,\thanks{email:~\href{mailto:dmitry.karlovets@metalab.ifmo.ru}{dmitry.karlovets@metalab.ifmo.ru}}}}

\affil{\textit{\textsuperscript{1} School of Physics and Engineering, ITMO University, 9 Lomonosova St., Saint-Petersburg, 191002, Russia}}
\affil{\textit{\textsuperscript{2} Joint Institute for Nuclear Research, 6 Joliot-Curie St., Dubna, 141980, Moscow region, Russia}}
\affil{\textit{\textsuperscript{3} Budker Institute of Nuclear Physics, Siberian Branch of RAS, 11 Lavrentiev Ave., Novosibirsk, Novosibirsk region, 630090, Russia}}
\affil{\textit{\textsuperscript{4} Institute of Nuclear Physics, 1 Ibragimov St., Almaty, 050032, Kazakhstan}}
\affil{\textit{\textsuperscript{5} Petersburg Nuclear Physics Institute of NRC ``Kurchatov Institute'', 1, mkr. Orlova roshcha, Gatchina, Leningradskaya Oblast, 188300, Russia}}

\begin{document}

\date{}
\maketitle

\begin{abstract}
    We present the design and an analysis of the classical and quantum beam dynamics for an S-band radio-frequency (RF) photoinjector dedicated to the generation of 
    relativistic vortex electron beams carrying a quantized orbital angular momentum (OAM). The field-map-based tracking simulations were performed in the ultra-low 
    bunch charge regime ($Q = 0.63$~pC), where space-charge effects are strongly suppressed. We demonstrate that optimizing the injection phase and solenoidal focusing preserves 
    the intrinsic thermal limit of the photoemission source, yielding a stable, emittance-compensated beam with a final kinetic energy of 4.9~MeV and a normalized transverse emittance of 
    $56\ \pi \cdot \mathrm{\mu m \cdot mrad}$. Furthermore, our quantum-mechanical analysis of 
    single-electron Laguerre--Gaussian wave packets inside the designed accelerating fields reveals that rapid energy gain suppresses transverse quantum spreading by nearly three 
    orders of magnitude compared to free-space propagation. This suppression protects the characteristic OAM-dependent spatial structure of the wave packet. These results establish 
    the beam-dynamics operating regime and vacuum conditions required for subsequent experiments on the generation and transport of relativistic vortex electron beams.
\end{abstract}

\maketitle

\section{\label{sec:Introduction}Introduction}

    The generation of \textit{relativistic vortex electron beams} -- quantum states with a helical phase front and a quantized orbital angular momentum projection (OAM) -- is attracting 
    growing interest across accelerator, atomic, and nuclear physics, electron microscopy, quantum optics, and foundation of quantum 
    mechanics~\cite{Knyazev18, Bliokh17, Karlovets21, Ivanov22}. To date, vortex electrons have been generated only in the sub-MeV range, up to kinetic energies of $\sim 300$~keV of 
    transmission electron microscopes~\cite{Uchida10, Verbeeck10, Blackburn14}. Extending such structured states into the multi-MeV regime relevant for photoinjectors and 
    linacs remains an outstanding challenge, yet promises significant impact. Specifically, transferring a well-defined topological charge
    -- an integer phase-winding number $\ell = 0, \pm 1, \pm 2, \ldots$ defined by the azimuthal phase factor $\exp(i\ell\varphi)$ of the electron wave function and 
    corresponding to an intrinsic OAM projection of $\ell\hbar$ per particle -- to relativistic beams opens pathways for novel beam-target interactions, 
    probing fundamental symmetries in hadronic systems, and developing advanced diagnostics for future colliders where controlling the internal degrees of freedom of the beam 
    is paramount~\cite{SchattschneiderPRB12, Wu19, Ivanov20, Ivanov20_2, Ivanov22, An25}.

    It is important to distinguish the genuine vortex beams of single particles whose wave functions are characterized by a helical phase structure $\exp(i\ell\varphi)$ 
    and a quantized OAM projection $\ell\hbar$~\cite{Bliokh17} from beams with modified transverse emittance ratios -- such as flat and round beams demonstrated at 
    Fermilab~\cite{Danilov96, Brinkmann02, Sun04, Piot06}. These latter beams can possess a so-called {\it extrinsic (non-quantized)} angular momentum, which is a property of the beam as a whole, 
    so a quantum state of each particle in the beam is not defined, whereas the former quantum beams of single electrons posses the so-called {\it intrinsic (quantized)} orbital angular momentum, 
    the property of each electron, complementary to spin~\cite{Bliokh17, Floettmann20, Karlovets21, Ivanov22}.

    One feasible route towards high-energy vortex electrons is the transfer of OAM from twisted ultraviolet light to electrons during 
    photoemission~\cite{Karlovets21, Pavlov24, Kazinski25}. Here, "structured light" specifically refers to spatially structured optical modes 
    (e.g., Laguerre--Gaussian modes) possessing a helical phase front and carrying orbital angular momentum~\cite{AndrewsBabiker12, Kotlyar22, Fu24}. This spatial structure is distinct from circular polarization, 
    which characterizes the spin angular momentum of the optical field. The generation of pure vortex states with a definite angular momentum projection and vanishing quantum 
    uncertainty is challenging due to finite temperature, space-charge effects, and other factors~\cite{Floettmann20, Karlovets21,Kazinski25}. However, the generation of a 
    superposition of vortex states with an OAM uncertainty smaller than the mean OAM value appears feasible for sufficiently high OAM (in our case up to $\ell=64$, corresponding to 
    an OAM projection of $64\hbar$), provided the bunch charge is kept low to mitigate space-charge effects.
    
    Realizing this mechanism requires a photoinjector capable of producing low-emittance beams with well-controlled transverse profiles and rapid acceleration, ensuring that the initial 
    OAM-imprinted phase structure of the emitted electron is preserved during the early acceleration. To this end, a dedicated S-band radio-frequency (RF) photoinjector test bench is being developed at 
    the Laboratory of Nuclear Problems (DNLP), Joint Institute for Nuclear Research (JINR), combining a 1.5-cell RF gun~\cite{Nikiforov18} and a high-power UV laser 
    system~\cite{Gacheva14}. Along with relativistic vortex electrons, there are plans at the  Institute of Modern Physics in China to put a twist on relativistic ions~\cite{An25}.

    RF photoinjectors are widely used as high-brightness electron sources at facilities such as LCLS~\cite{Alley99}, the European XFEL~\cite{Ferrario01}, and SuperKEKB~\cite{Cahill14}, 
    owing to their ability to quickly suppress space-charge effects via high-gradient acceleration~\cite{Giribono23, Maltseva24}. At JINR, there exists a LINAC facility, which is a 
    linear accelerator of electrons to the energy of 200 MeV and plans to come to 400 MeV in 2027~\cite{Nozdrin25}, it operates with a thermionic DC gun and a multi-stage RF bunching 
    system~\cite{Nozdrin21}, producing $\sim 1$--$2$~ps bunches. Replacing it with an RF photoinjector enables direct generation of 10~ps bunches (FWHM) in single-bunch operation, 
    with the significantly improved transverse beam quality and precise control over the charge and transverse profile -- critical prerequisites for producing structured quantum states.
    
    Before addressing the full complexity of macroscopic vortex beam generation, which inherently includes thermal emittance, stochastic photoemission, and space-charge 
    decoherence, it is necessary to establish the fundamental limits of the acceleration process itself. In this paper, we focus strictly on the transport and acceleration 
    mechanics to define a theoretical upper bound on the OAM preservation. This idealized single-particle baseline is a necessary prerequisite for designing and interpreting 
    future experiments at the test bench.
    
    In this work, we present the design of the JINR RF photoinjector test bench, specifically configured for the generation and acceleration of relativistic vortex 
    electron beams. To evaluate the proposed hardware layout, we perform field-map-based beam-dynamics simulations in the nominal ultra-low charge regime. These 
    simulations validate the chosen configuration, demonstrating that the integrated compensation solenoid provides effective transverse-envelope control. While the 
    normalized emittance remains defined by the intrinsic thermal limit of the photoemission source, the optimized design stabilizes transverse phase-space evolution 
    and strongly suppresses RF-induced distortions, thereby establishing the robust, reproducible beam-dynamics conditions required for subsequent experiments on vortex beam generation.
    
    We also analyze the quantum evolution of single-electron Laguerre--Gaussian wave packets with OAM values up to $\ell = 64$ in the same accelerating fields. The transverse 
    quantum spreading is found to be strongly suppressed compared to free-space propagation due to the rapid increase in longitudinal momentum, an effect whose physical origin 
    was revealed and explained in Ref.~\cite{z6j7-grs2} and confirmed in Ref.~\cite{FilinaBaturin2026}, allowing realistic estimates of 
    the transverse coherence length under experimentally accessible conditions. Notably, vortex beams beyond the paraxial limit (where $p_\perp$ is of the order of $p_z$) are 
    inapplicable for experimental use specifically due to transverse quantum spreading, except for the case of their prompt acceleration~\cite{z6j7-grs2, FilinaBaturin2026}. 
    The sensitivity of this baseline to experimentally relevant imperfections is assessed using a 
    phenomenological incoherent-ensemble model for the nominal low-charge regime, incorporating the energy-spread evolution obtained from beam-dynamics simulations, as well as 
    RF amplitude and phase jitter, transverse source offsets, and variations of the solenoidal focusing field.
    
    Within the experimentally relevant parameter ranges considered here, these effects lead to a reduction of the expected ring contrast 
    (as quantified below from the transverse probability-density maps in Sec.~\ref{sec:quantum}) and an increase of the effective transverse 
    width, but do not destroy the characteristic vortex structure. This supports the robustness of the proposed scheme for generating high-OAM electron beams under realistic 
    operating conditions.

\section{\label{sec:test_bench_design}Test bench design}

\subsection{\label{sec:rf-gun}RF photogun}

\begin{figure}[!ht]
        \centering
        \includegraphics[width=0.75\linewidth]{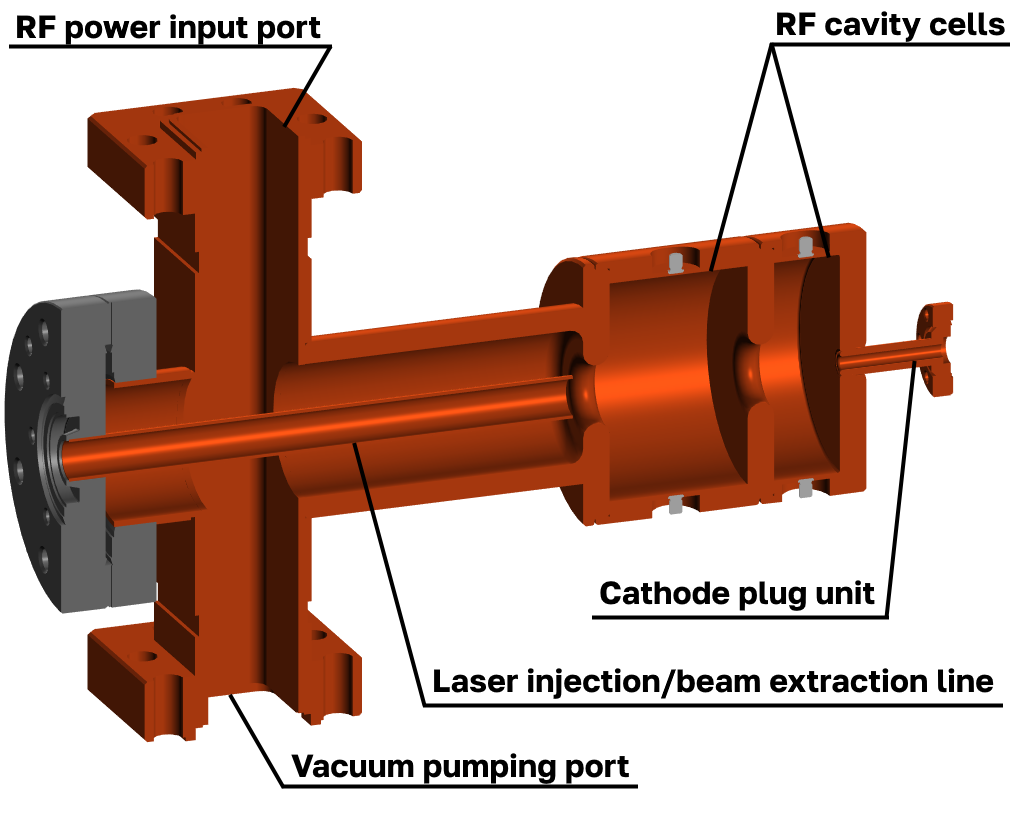}
        \caption{Internal structure of the 1.5-cell S-band RF photogun operating in the $\pi$-mode (2856~MHz). The central aperture is used for both UV drive laser injection and electron 
        beam extraction.}
        \label{fig:rf_gun}
    \end{figure}

\begin{figure}[!ht]
        \centering
        \includegraphics[width=0.75\linewidth]{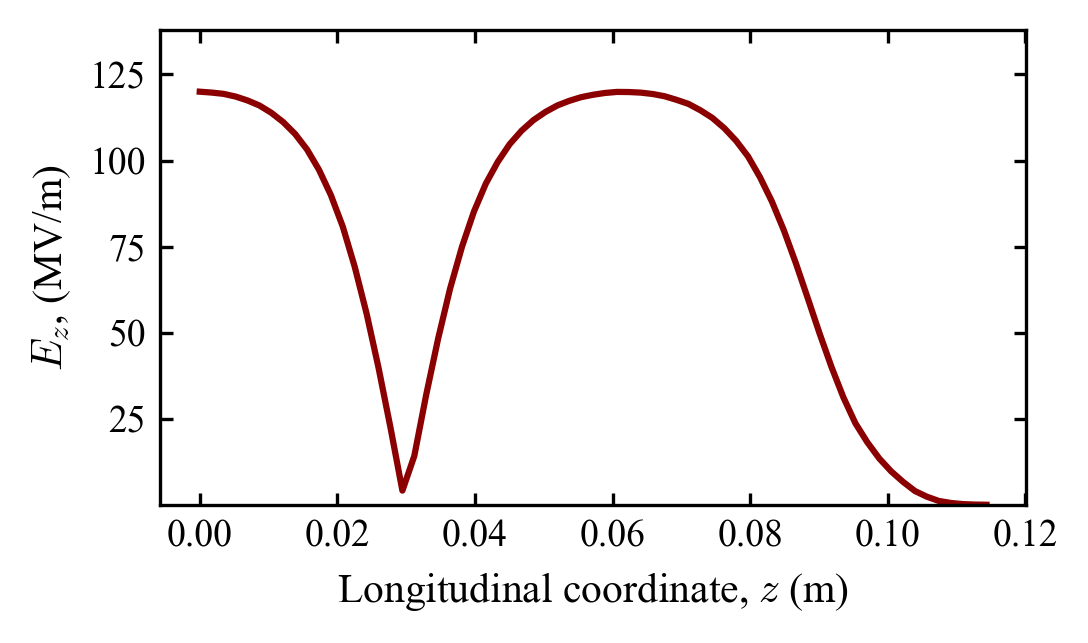}
        \caption{On-axis accelerating electric field profile $E_z(z)$ along the RF photogun cavity, obtained from CST Studio Suite eigenmode simulations~\cite{Samofalova21}, corresponding to an input 
        RF power of 10~MW.}
        \label{fig:efield}
    \end{figure}

    The core component of the RF photoinjector test bench is a 1.5-cell RF photogun developed and manufactured at the Budker Institute of Nuclear Physics 
    (BINP SB RAS)~\cite{Nikiforov18}. The eigenfrequency of the gun is 2856~MHz with a $\pi$ operating mode~\cite{Samofalova21}. Figures~\ref{fig:rf_gun} 
    and~\ref{fig:efield} show the internal structure of the gun and the accelerating field gradient simulated using the CST Studio suite~\cite{CST}. The experimentally 
    measured voltage standing wave ratio (VSWR) is 1.13~\cite{Nozdrin26}, and the design unloaded quality factor of the cavity is 12000.
    
    Pure oxygen-free copper was chosen as the photocathode material. A copper insert was installed in the cathode assembly flush with the back wall of the photogun. 
    The selection of this cathode type is primarily driven by its inherent operational simplicity compared to semiconductor photocathodes, such as $\mathrm{Cs_2Te}$. Despite a 
    relatively low quantum efficiency ($\mathrm{QE} \approx 10^{-5}$~\cite{Xiang15}), copper exhibits a fast emission response time (on the order of 100~fs~\cite{Loisch22}). Furthermore, copper does not 
    impose the stringent ultra-high vacuum requirements associated with semiconductors.

\subsection{\label{sec:rf_power_supply}RF power supply system}

    The RF power supply system of the photogun cavity is designed identically to the system used in the DLNP LINAC electron linear accelerator~\cite{Nozdrin21}. The core of the entire 
    power system is a Thomson TH~2129 pulsed klystron. It features a maximum output power of 20~MW at an operating frequency of 2856~MHz, and an RF power envelope duration ranging 
    from 1 to 6~$\mu$s. Although the system is capable of a repetition rate of up to 10~Hz, our current standard operating regime is restricted to 5~Hz. This limitation is dictated by 
    the capacity of the existing klystron cooling system, which cannot sustain prolonged continuous operation at the maximum repetition rate and requires future modernization.

    An AKIP-7SG384 RF signal generator (equivalent to the Stanford Research Systems SG384) serves as the master oscillator, providing an initial output power of 10~mW. 
    This RF signal is subsequently pre-amplified to a nominal operating power of 100~W (with a maximum capability of 150~W) and fed into the klystron via a coaxial cable. From the klystron output, the amplified high-power RF wave is transported 
    to the photoinjector cavity through a waveguide line. During the calibration of the RF system, it was determined that with the pulse-forming network (PFN) operating across all 40 
    of its constituent modules~\cite{Kobets2014}, the maximum klystron output power currently reaches 10~MW. This input power corresponds to an accelerating field gradient of 
    approximately 120~MV/m within the photogun cavity.

    \begin{figure}[!ht]
        \centering
        \includegraphics[width=0.75\linewidth]{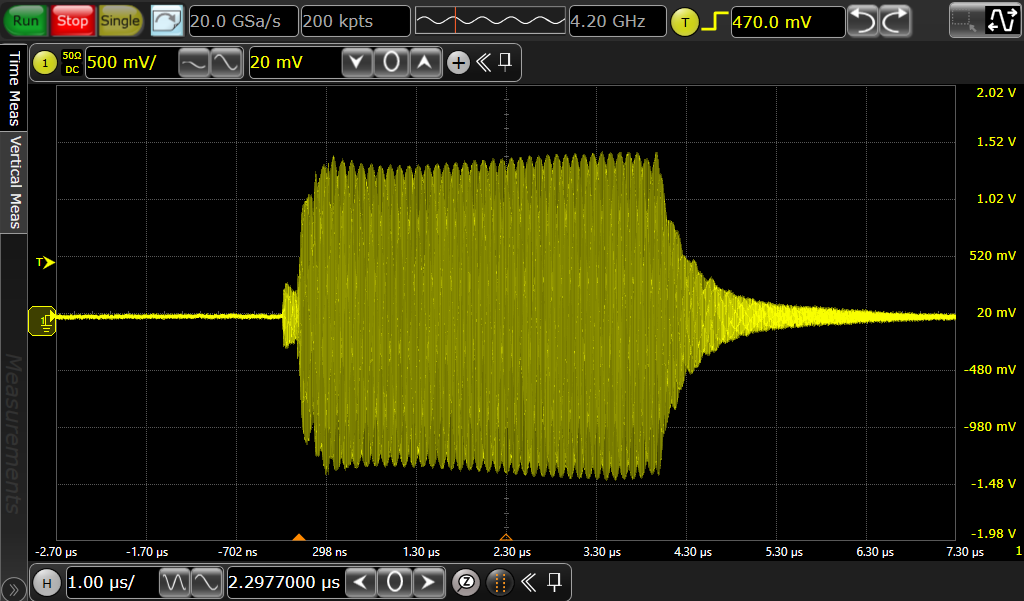}
        \caption{Oscillogram of the forward RF pulse envelope. The recorded amplitude of 1.42~V at the oscilloscope input corresponds to an evaluated klystron 
        output power of 10~MW.}
        \label{fig:rf_forward}
    \end{figure}
    
    The successful delivery of this power is demonstrated by the RF pulse envelope oscillogram shown in Fig.~\ref{fig:rf_forward}. The RF power calibration was 
    performed using a diagnostic line comprising a 50~dB forward directional coupler, 30~dB of coaxial attenuators, and signal cables introducing an additional 7~dB of loss, yielding 
    a total attenuation of 87~dB. Under maximum klystron excitation, the detected signal amplitude at the 50~$\Omega$ oscilloscope input was 1.42~V. This corresponds to a measured RF 
    peak power of $+13$~dBm. Accounting for the total line attenuation, the evaluated maximum klystron output power is 100~dBm, or 10~MW.
    
\subsection{\label{sec:uv_drive_laser}UV drive laser}
    
    The UV drive laser is based on a mode-locked ytterbium-doped fiber oscillator at $\lambda = 1047$~nm with a SESAM~\cite{Keller96} for pulse shortening; it produces pulses with 
    the duration of $\tau = 10\ \mathrm{ps}$ (FWHM)~\cite{Gacheva14}, which corresponds to an rms value of 4.225~ps. Two lamp-pumped Nd:YLF amplifiers raise 
    the pulse energy before sequential generation of the second (KTP) and fourth (BBO) harmonics. 
    
\begin{figure}[!ht]
    \centering
    \includegraphics[width=0.75\textwidth]{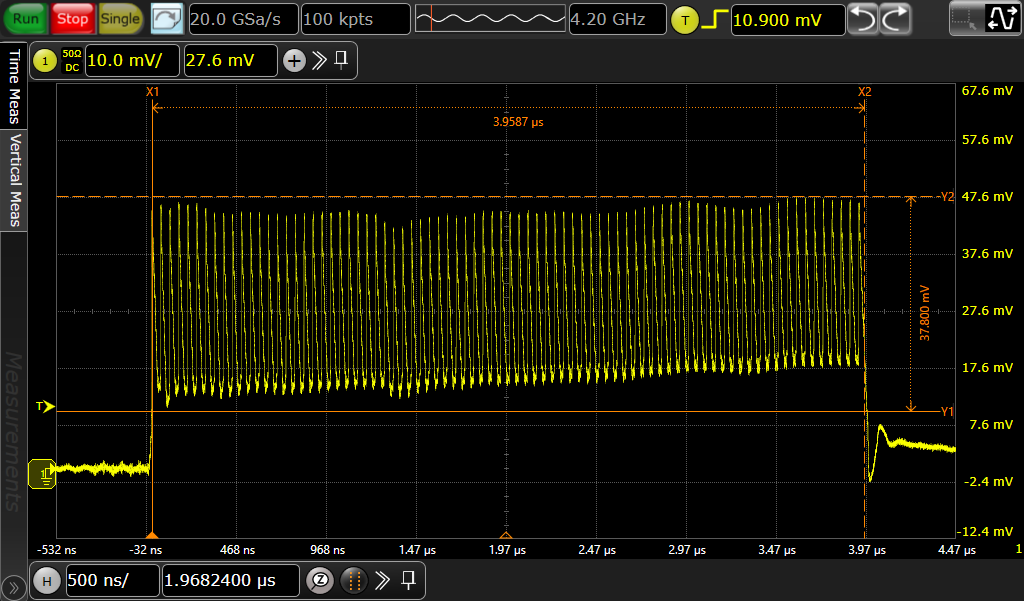}
    \caption{Oscillogram of the generated UV drive laser pulse train. The signal demonstrates a tailored sequence of 80 micropulses formed by the modernized 
    acousto-optic modulator (AOM). The total duration of the train is approximately 3.96~$\mu$s, allowing it to fit perfectly within the RF pulse envelope.}
    \label{fig:aom_train}
\end{figure}

    A recent upgrade~\cite{Dyatlov24} increased the repetition rate to the $71^\mathrm{st}$ subharmonic of the RF frequency (40.225~MHz) and introduced synchronized amplifier 
    control, improving UV pulse stability and achieving 3--4~$\mu$J energies per pulse. Furthermore, the pulse-picking system, which employs an acousto-optic 
    modulator (AOM), was modernized. While the previous setup was limited to forming fixed trains of exactly 8000 pulses with a 100~ns inter-pulse spacing, the upgraded system 
    offers tunable train sizes ranging from 25 to 16,000 pulses, with a minimum pulse spacing of 50~ns. This newfound flexibility makes it possible to generate a tailored train of 
    70--80 pulses with a total train duration of 3--4~$\mu$s (see Fig.~\ref{fig:aom_train}), allowing it to fit perfectly within the RF envelope.

\subsection{\label{sec:laser_transport}Laser transport line}

\begin{figure*}[!ht]
        \centering
        \includegraphics[width=0.95\textwidth]{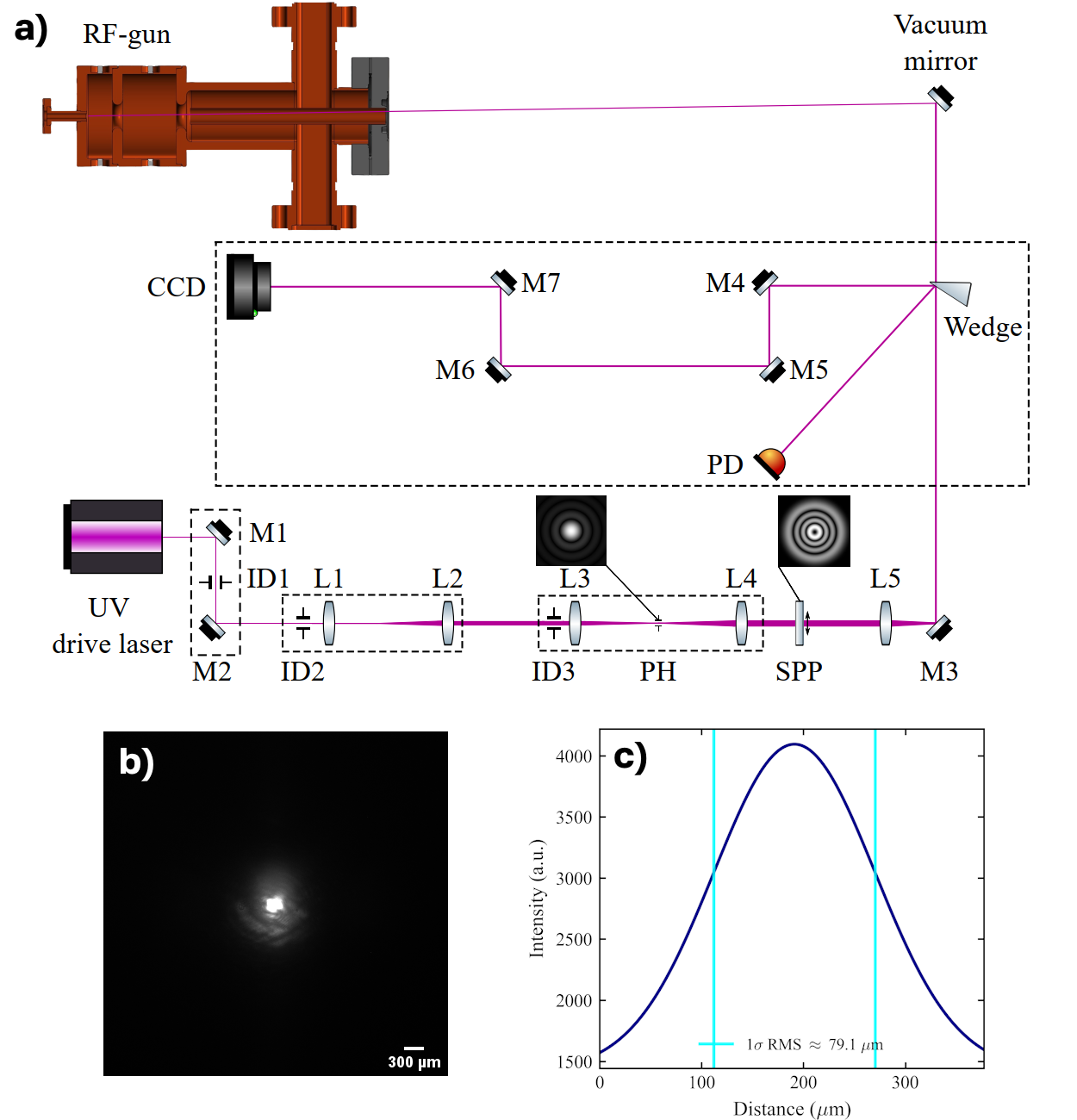}
        \caption{Schematic of the UV drive laser transport and diagnostic system. 
        (a) The UV drive laser beam is directed by a periscope (mirrors M1 and M2) and passes through an iris diaphragm (ID1). It then enters a collimating telescope consisting of 
        lenses L1 and L2 with an intermediate iris diaphragm (ID2). The expanded beam is cleaned by a spatial filter comprising an iris diaphragm (ID3), lens L3, a pinhole (PH), and 
        lens L4. The beam then traverses a spiral phase plate (SPP) and a focusing lens (L5), and is steered by mirror M3 through an optical wedge. The SPP is an optional element 
        used exclusively in experiments with twisted photons. The main transmitted beam is reflected by an in-vacuum mirror onto the RF-gun photocathode. The wedge picks off two 
        reflections for diagnostics: one is directed to a photodiode (PD), and the other is routed via mirrors M4, M5, M6, and M7 to a CCD camera.
        (b) Measured laser beam spot profile on a virtual photocathode, 
        (c) Transverse intensity profile (blue curve) taken across the spot center, extracted from panel (b). Cyan vertical lines indicate the 1$\sigma$ root-mean-square (RMS) width 
        of approximately 79.1 $\mu$m.}
        \label{fig:laser_transport}
    \end{figure*}

    The layout of the laser beam transport line to the photocathode and its key components are presented in Fig.~\ref{fig:laser_transport}(a). The UV drive laser and the accelerating 
    structure are housed in separate rooms to ensure radiation safety and maintain cleanroom conditions. An opening in the separating wall connects the two areas. This opening 
    is equipped with a UV fused silica window that transmits up to 95\% of the UV radiation. A periscope (M1-M2) transfers the beam between the rooms and elevates the transport 
    line to the RF photoinjector axis. Furthermore, dovetail optical rails are installed along the walls to facilitate the beam routing.

    A collimating telescope (L1-L2) is mounted on the rail in the UV drive laser room. It uses lenses with focal lengths of 223~mm and 300~mm. This telescope initially 
    restores the beam wavefront and expands the beam to a diameter of approximately 5~mm. A spatial filter is installed on the rail in the accelerating structure room. It 
    consists of lens L3 ($f = 125$~mm), a pinhole PH ($d = 100~\mu$m), and lens L4 ($f = 240$~mm). This spatial filter is essential because the UV beam is generated via nonlinear 
    frequency quadrupling using BBO crystal. The nonlinear conversion in the BBO crystals introduces phase distortions and high-order aberrations. 
    These effects partially degrade the wavefront of the output UV beam. The spatial filter eliminates these artifacts and cleans the transverse mode profile.

    Further downstream, a spiral phase plate (SPP) with a topological charge of $\ell = 64$ is installed. As detailed in Ref.~\cite{Dyatlov26}, this diffractive 
    optical element generates a pure Laguerre--Gaussian mode carrying an orbital angular momentum of $64\hbar$. The resulting ring-shaped laser beam exhibits a FWHM of $800\ \mu\mathrm{m}$. For such a high-order spatial mode, this macro-profile corresponds to an rms transverse size of $\sigma_{x,y} \approx 280\ \mu\mathrm{m}$. 
    To compress this structured beam down to the target rms size of $\sim 100\ \mu\mathrm{m}$ required for optimal operation of the photoinjector, an additional short-focus optical 
    telescope must be integrated into the transport line immediately after the SPP.

    The utilization of such a high-OAM optical drive is highly advantageous for the photoemission process. Imprinting a large topological charge onto the electron wave 
    packets maximizes the spectral separation between the target vortex state and any spurious low-OAM components. This separation is critically important to mitigate the finite OAM 
    bandwidth inevitably introduced by the bounded transverse size of the emission area and residual spatial misalignments at the cathode. By operating at a significantly elevated OAM 
    quantum number, the experimental distinguishability of the generated vortex electron beam is substantially improved, ensuring that its macroscopic topological properties remain 
    robust against initial emission imperfectiosns.

    It is important to note that $\ell = 64$ is not an upper limit for our photoemission scheme, which can also generate vortex electrons with 
    $\ell > 100$. During photoemission, the resulting electron OAM bandwidth is governed primarily by the topological charge $\ell$ of the incident 
    UV laser beam~\cite{Shchepkin2026}. Fundamentally, there is no strict upper bound on the topological charge of optical vortices; for instance, Fickler \textit{et al.} 
    experimentally generated twisted photons with $\ell = 10010$~\cite{Fickler16}. In our setup, reaching higher electron OAM values simply requires 
    fabricating a more complex diffractive optical element and adapting the telescope optics to match the beam spot size on the photocathode.

    Next, lens L5 ($f = 2500$~mm) focuses the laser beam, and mirror M3 steers it toward the copper photocathode. An in-vacuum mirror delivers the radiation directly 
    to the photocathode surface. This mirror features a highly polished, oxygen-free copper substrate. A UV-reflective coating was deposited onto it via thermal evaporation. The 
    coating consists of a 50~nm chromium adhesion sublayer and a 200~nm aluminum top layer. This technique ensured a mirror surface of near-diffraction-limited quality.

    A UV fused silica wedge picks off approximately 10\% of the laser radiation. This produces two separate reflections, each carrying $\approx 5\%$ of the initial 
    intensity. The first reflection goes to a fast UV photodiode (FDS010, Thorlabs). This diode monitors the pulse train shape generated by the drive laser. The second reflection goes to an 
    SDU-285R CCD camera (SpetsTeleTekhnika). The camera was modified for UV sensitivity by decapping the sensor and removing its protective glass. Mirrors M4 -- M7 adjust the optical path length to 
    match the 2500~mm focal length.

    This entire optical system enables the formation of a laser beam with an rms size of approximately 79~$\mu$m on the cathode 
    (Figs.~\ref{fig:laser_transport}(b) and \ref{fig:laser_transport}(c)). It should be noted that high spatial coherence of the initial beam is crucial for experiments involving 
    twisted photons. While the achieved spot size is a satisfactory result for the current stage, further upgrades to the optical alignment system will enable a tighter focus on 
    the cathode, which in turn will enhance the spatial coherence of the beam.  

\subsection{\label{sec:emittance_compensation}Emittance compensation solenoid}

\begin{figure}[!ht]
        \centering
        \includegraphics[width=0.75\linewidth]{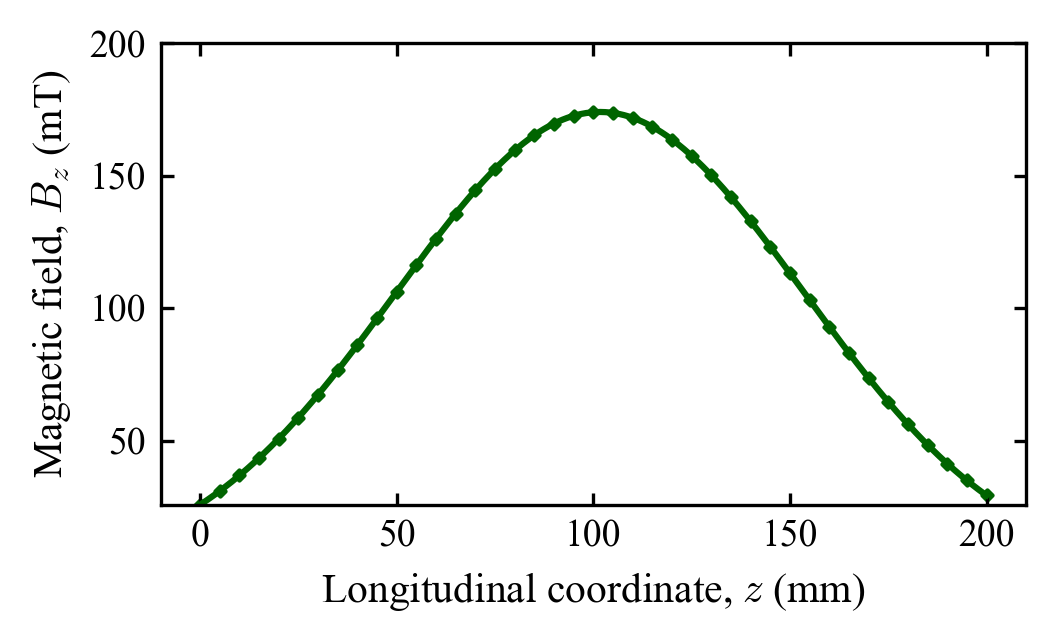}
        \caption{Measured on-axis magnetic field profile $B_z(z)$ of the emittance compensation solenoid (5~mm longitudinal sampling).}
        \label{fig:solenoid_field}
    \end{figure}

    The emittance compensation solenoid is primarily designed to mitigate the transverse projected emittance growth caused by intense space-charge forces near the photocathode. By 
    focusing the expanding electron beam, its magnetic field realigns the phase space ellipses of different longitudinal slices, effectively minimizing the overall projected 
    emittance along the beamline~\cite{Carlsten89}. The prototype of this solenoid was manufactured at JINR. It has an inner radius of 100~mm, outer radius of 140~mm, a mechanical length of 100~mm, and 
    contains 216 copper turns arranged in six layers. The axial magnetic-field profile was measured with a longitudinal step of 5~mm. The measured distribution is shown in 
    Fig.~\ref{fig:solenoid_field}. Because the discrete measurement resolution and experimental constraints make it difficult to resolve a perfectly smooth field decay toward 
    the edges, the measured data were numerically interpolated and slightly smoothed to obtain a continuous field profile suitable for beam-dynamics simulations. The resulting 
    profile preserves the measured peak field and overall integral strength, while providing a physically consistent description of the fringe fields extending beyond the mechanical 
    length of the coil.

\subsection{\label{sec:focusing_solenoid}Focusing solenoid}

    The focusing solenoid used in the test bench primarily serves to deliver the photogun-generated electron beam onto the target. Furthermore, 
    during the initial experimental phases, it will be employed for beam emittance measurements using the solenoid scan technique~\cite{Dechoudhury22}. This method relies on measuring the beam 
    waist variation as a function of small changes in the magnetic field induction, with the transverse beam profile recorded on a scintillation screen (in our case, a Ce:YAG 
    scintillator). The dependence of the squared rms beam size, $\sigma_{x,y}^2$, on the solenoid focusing strength is fitted using the least-squares method. The elements of the 
    initial beam matrix ($\sigma_{11}$, $\sigma_{12}$, $\sigma_{22}$) are extracted from this parabolic fit and subsequently used to calculate the geometric beam emittance.

\begin{figure}[!ht]
        \centering
        \includegraphics[width=0.75\linewidth]{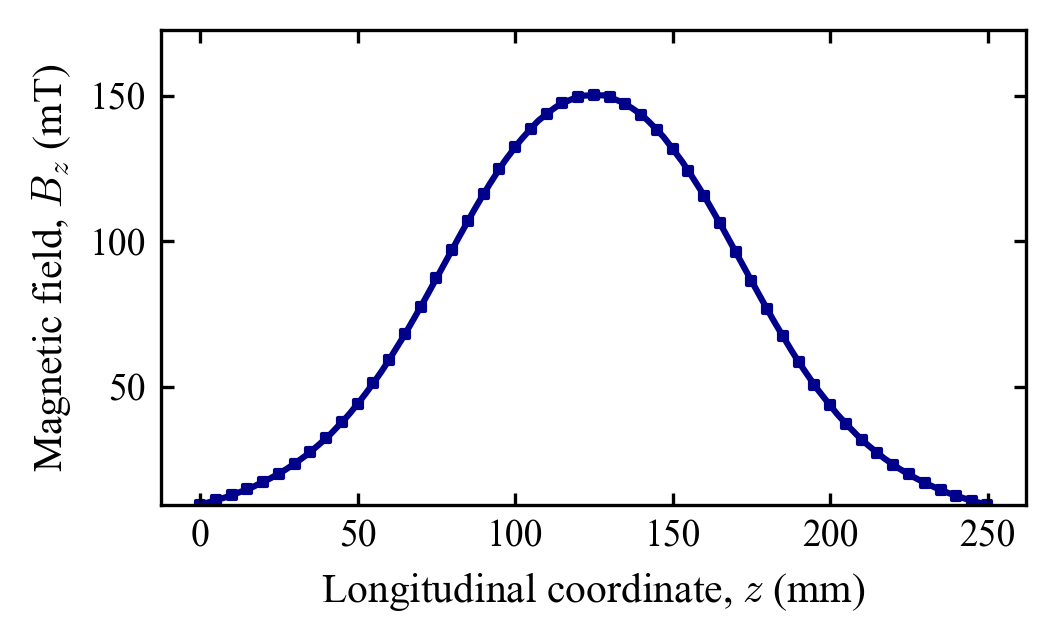}
        \caption{Measured on-axis magnetic field profile $B_z(z)$ of the focusing solenoid (5~mm longitudinal sampling).}
        \label{fig:focusing_solenoid_field}
    \end{figure}

    For beam focusing, we utilize an existing solenoid with an inner radius of 129~mm and a mechanical length of 105~mm. The on-axis magnetic field profile of this 
    focusing solenoid is presented in Fig.~\ref{fig:focusing_solenoid_field}. The field mapping procedure was performed identically to that of the emittance-compensation solenoid: a 
    Hall-effect magnetic field probe was moved along the axis in steps of approximately 5~mm on a dedicated test stand.

\subsection{\label{sec:steering_magnet}Steering magnet}

    The steering magnet is designed for electron beam orbit correction. Initial axis alignment is established via the laser transport system, where the overall alignment precision 
    inherently depends on the accurate delivery of the laser spot to the center of the photocathode. Additionally, an auxiliary transverse magnetic field component, $H_y$, which 
    arises within the photogun cavity, assists in centering the beam along its nominal trajectory.

    The yoke of the corrector magnet is fabricated from unlaminated steel. At maximum excitation, the sextupole component of the magnetic field does not 
    exceed $0.015\ \mathrm{m}^{-2}$. The coils generate a maximum magnetic field of 15~mT at an operating current of 3.5~A. The physical dimensions of the magnet include a yoke 
    length of 140~mm and an aperture of $146 \times 132\ \mathrm{mm}$. It should be noted that initial beam axis misalignment is neglected in the subsequent beam dynamics 
    simulations; consequently, the contribution of the steering magnet is not considered in the computational model.

\subsection{\label{sec:quadrupole_doublet}Quadrupole doublet}

    The primary purpose of the quadrupole doublet is to perform beam emittance measurements using the quadrupole scan technique~\cite{Mostacci12}. Conceptually, this method is similar to the 
    solenoid scan; however, they exhibit fundamental differences in terms of beam dynamics. A solenoid not only focuses the beam but also introduces a Larmor rotation in the transverse 
    plane around the propagation axis, inherently coupling the horizontal and vertical beam motion. To accurately measure the emittance using a solenoid, this rotation must be 
    rigorously accounted for in the matrix formalism. In contrast, a quadrupole acts independently in the two orthogonal transverse planes (focusing in one plane while defocusing in 
    the other). This decoupling significantly simplifies the experimental data processing and improves the reliability of the emittance reconstruction.

    The second crucial function of this doublet is the diagnostic analysis of the generated Laguerre-Gaussian (LG) vortex electron beam using the induced astigmatism 
    method~\cite{SchattschneiderPRL12}. When a vortex beam propagates through an electron-optical element with different focal strengths in the two orthogonal planes (such as a 
    quadrupole), its axial symmetry is broken, and the LG modes are converted into asymmetric Hermite-Gaussian (HG) modes. The OAM value, $\ell$, of the 
    initial beam can be determined by analyzing the resulting HG mode order, specifically by counting the number of transverse nodal lines (or intensity lobes) in the beam profile. 
    Furthermore, the tilt direction of the converted HG pattern unambiguously indicates the sign of the initial LG mode topological charge.

    Currently, this focusing cascade is in the early stages of development. The quadrupole magnets being designed feature an aperture of 40~mm and an effective 
    magnetic length of approximately 140~mm.

\subsection{\label{sec:dipole_magnet}Dipole magnet}

\begin{figure}[!ht]
        \centering
        \includegraphics[width=0.75\linewidth]{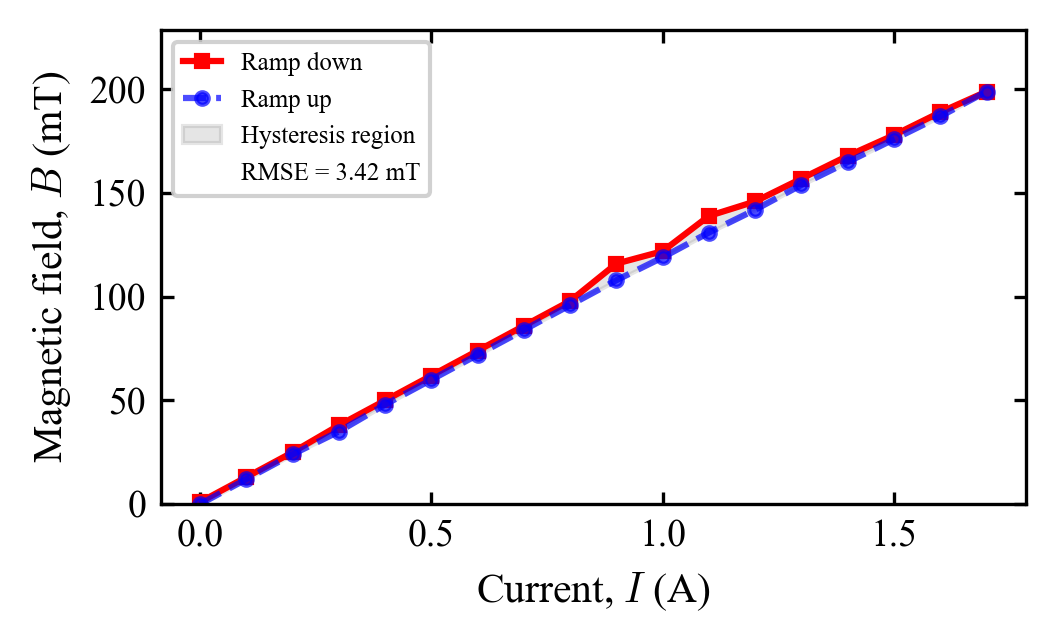}
        \caption{Measured magnetic field induction $B$ as a function of the excitation current $I$ for the sector dipole magnet. The plot illustrates the hysteresis behavior, displaying 
        both the ramp-up (blue circles, dashed line) and ramp-down (red squares, solid line) excitation cycles. The shaded area highlights the hysteresis region, characterized by a 
        root-mean-square error (RMSE) of 3.42~mT between the two branches.}
        \label{fig:dipole_field}
    \end{figure}

    The dipole magnet is designed to measure the output electron beam energy in the photoinjector test bench beamline. This device is configured as a sector 
    dipole with a bending angle of $28.5^\circ$ and a bending radius of 550~mm. The magnetic field induction as a function of the excitation current is presented in Fig.~\ref{fig:dipole_field}. Based 
    on the expected relativistic particle momentum ($p \approx 5.49$~MeV/$c$) and the corresponding magnetic rigidity of the beam ($B\rho \approx 0.018$~T$\cdot$m), the required 
    homogeneous magnetic field induction to bend the beam by the specified angle is approximately 33~mT. The obtained field dependence demonstrates that the capability of this dipole 
    is fully sufficient for resolving the complete energy spectrum of the beam.

\subsection{\label{sec:complete_test_bench_layout}Complete test bench layout}

\begin{figure}[!ht]
        \centering
        \includegraphics[width=1\linewidth]{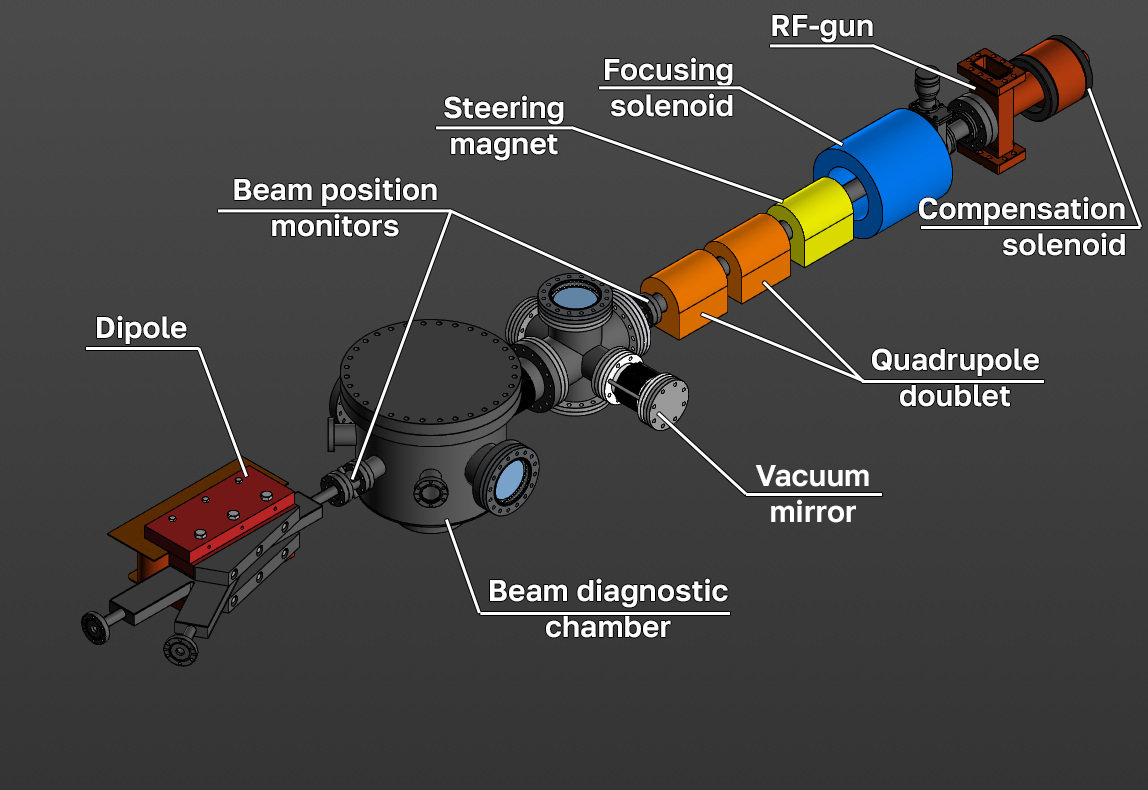}
        \caption{3D CAD model of the complete DLNP JINR RF photoinjector test bench.}
        \label{fig:test_bench}
    \end{figure}

    The initial commissioning of the test bench (Phase 1) was carried out using a preliminary configuration of the setup~\cite{Nozdrin26}. In this work, we consider 
    the upgraded configuration (Phase 2), shown in Fig.~\ref{fig:test_bench}. The total length of the beamline in this configuration is 2.9~m. Vacuum pumping is provided by Agilent 
    VacIon Plus 75 ion pumps with a nominal pumping speed of 75~l/s. These pumps maintain an operating pressure of $10^{-9}$~Torr in the photogun region and $10^{-8}$~Torr throughout 
    the remainder of the vacuum system.

    Furthermore, in the final Phase 2 design, the existing emittance-compensation and focusing solenoids will be replaced with water-cooled versions utilizing hollow copper conductors 
    that simultaneously carry the cooling water and the excitation current. This upgrade is dictated by the necessity of 24-hour continuous operation, whereas the current uncooled 
    coils can only sustain nominal magnetic fields for a relatively short duration due to thermal limitations. Although the overall physical dimensions of the upgraded solenoids will 
    increase to accommodate the cooling system, their effective magnetic lengths and the on-axis magnetic field profiles will be strictly preserved to maintain the validated beam 
    dynamics. Nevertheless, for the initial proof-of-principle experiments, the operational time window provided by the current uncooled coils is entirely sufficient.

    During the initial experimental phase, only a subset of the depicted components will be utilized: the emittance-compensation and focusing solenoids, the steering magnet, the 
    diagnostic chamber, and the dipole magnet. This staged approach is necessitated by the fact that the quadrupole doublet and the beam position monitors (BPMs) are currently in 
    early stages of development and will not be integrated during the initial commissioning of the test bench. In the absence of BPMs, primary beam trajectory alignment will be performed 
    using a Ce:YAG scintillation screen located within the diagnostic chamber. The scintillator is mounted on an in-vacuum manipulator providing two degrees of freedom 
    (linear translation $X$ and rotation $\theta$). This manipulator ensures a positioning precision of up to 10~$\mu$m along the $X$-axis and $1^\circ$ in the azimuthal 
    angle $\theta$.

    The defining feature of the assembled test bench is the integration of structured light directly at the electron emission source within a high-gradient 
    RF environment. Conventional methods for generating 200--300~keV vortex electron beams in transmission electron microscopes rely on passive material elements, such as amplitude 
    holograms or phase plates, placed directly in the electron beamline~\cite{Uchida10, Verbeeck10, Blackburn14}. At multi-MeV energies, the use of such invasive in-vacuum optics is 
    precluded by severe bremsstrahlung, multiple Coulomb scattering, and resultant beam degradation. The presented layout circumvents these limitations by utilizing the 
    Laguerre--Gaussian UV laser pulse to imprint the topological charge onto the electron wave 
    packets during photoemission. Immediately following emission, the 120~MV/m accelerating gradient of the photogun cavity provides rapid beam capture. The rapid 
    increase in longitudinal momentum strongly reduces the subsequent transverse evolution of the emitted wave packet and simultaneously suppresses collective space-charge effects 
    during the early acceleration stage. Ultimately, this hardware configuration provides a non-invasive pathway for generating and transporting structured quantum electron 
    states at relativistic energies.

    Within the standalone 1.5-cell RF photogun configuration, the kinetic energy of $4.9$~MeV represents a practical upper limit. The maximum accelerating 
    gradient ($\approx 120$~MV/m at $10$~MW input RF power) is constrained by the cavity geometry, its unloaded quality factor ($Q_0 = 12000$~\cite{Nikiforov18}), 
    and surface RF breakdown limits. Reaching beam energies of $10$~MeV and higher requires either replacing the gun with a multi-cell RF structure 
    (e.g., a $2.5$- or $3.5$-cell cavity) or integrating the photoinjector into the beamline of a linear accelerator. In particular, integrating the 
    photoinjector into the existing JINR LINAC facility beamline will enable the acceleration of vortex electron beams up to $200$~MeV~\cite{Nozdrin21, Nozdrin25}. 
    Notably, a recent study~\cite{Murtazin2026} demonstrates that although a vortex electron beam experiences gradual OAM loss in an accelerating field, 
    the characteristic timescale of this process is much longer than the beam transit time through the accelerator. Consequently, within a linear 
    accelerator, the upper energy limit is determined solely by the machine length and the energy gain rate rather than by OAM loss during acceleration.    

\section{\label{sec:classical_beam_dynamics}Classical beam dynamics and operating regime}

\subsection{\label{sec:beamline_dynamics}Photoinjector beamline dynamics and emittance compensation}

    Beam dynamics simulations were carried out with \mbox{ASTRA} \cite{ASTRA}, which combines a 3D particle-in-cell space-charge solver 
    (in cylindrical symmetry) with high-order Runge–Kutta tracking in external RF and magnetic fields. The principal input parameters are the 
    bunch charge (determined by photocathode quantum efficiency and laser pulse energy), the mean initial energy and its spread, the 
    normalized transverse emittance, the initial spatial and momentum distributions, and the RF phase at photoemission.
    
    The initial beam parameters are derived from the properties of the UV drive laser and the copper photocathode. The fourth-harmonic laser operates at $\lambda = 262\ \mathrm{nm}$, 
    corresponding to a photon energy of $4.74\ \mathrm{eV}$. With typical UV pulse energies of $3~\mu\mathrm{J}$ delivered to the vacuum chamber and taking into account the 
    measured transport losses, the number of photons incident on the cathode is of order $3.95\times 10^{11}$ per pulse.

    Using a copper photocathode with a quantum efficiency of $QE \sim 10^{-5}$~\cite{Xiang15}, this yields an emitted charge of approximately
    \[
        Q = 0.63\ \mathrm{pC},
    \]
    which is used as the baseline value in the simulations.

    The photoemission model follows the Dowell–Schmerge approach~\cite{Dowell09}, assuming Fermi–Dirac statistics and Gaussian spatial distributions. For an effective copper work 
    function of $\phi_{\rm eff}=4.3\ \mathrm{eV}$~\cite{Karkare15}, the resulting mean kinetic energy and energy spread of emitted electrons lie in the sub-eV range, consistent with 
    standard metallic photocathodes. With the UV laser rms beam size set to $\sigma_{x,y}=79\ \mathrm{\mu m}$ (equal to the initial RMS beam size), the corresponding intrinsic 
    normalized emittance is
\[
    \varepsilon_{n,\,x,y} = 46\ \pi \cdot \mathrm{\mu m\cdot mrad},
\]
    which defines the initial phase-space conditions used for the beam-dynamics simulations.

    To account for realistic transport, apertures corresponding to the internal RF-gun geometry were included (Fig.~\ref{fig:rf_gun}). 
    Without an external solenoidal field, most particles are lost within the first 25 cm from the cathode inside the coaxial input structure. Therefore,
    all simulations in the main paper body are conducted with the external solenoid field.
    
    At the present low bunch charge ($0.63\ \mathrm{pC}$), collective space-charge effects are significant only within the first few 
    millimeters after emission and rapidly diminish as the beam becomes relativistic. Consequently, the transverse emittance evolution is 
    governed primarily by RF-induced correlations and by magnetization effects associated with the solenoidal field, rather than by Coulomb 
    forces.

\begin{figure*}
        \centering
        \includegraphics[width=1\textwidth]{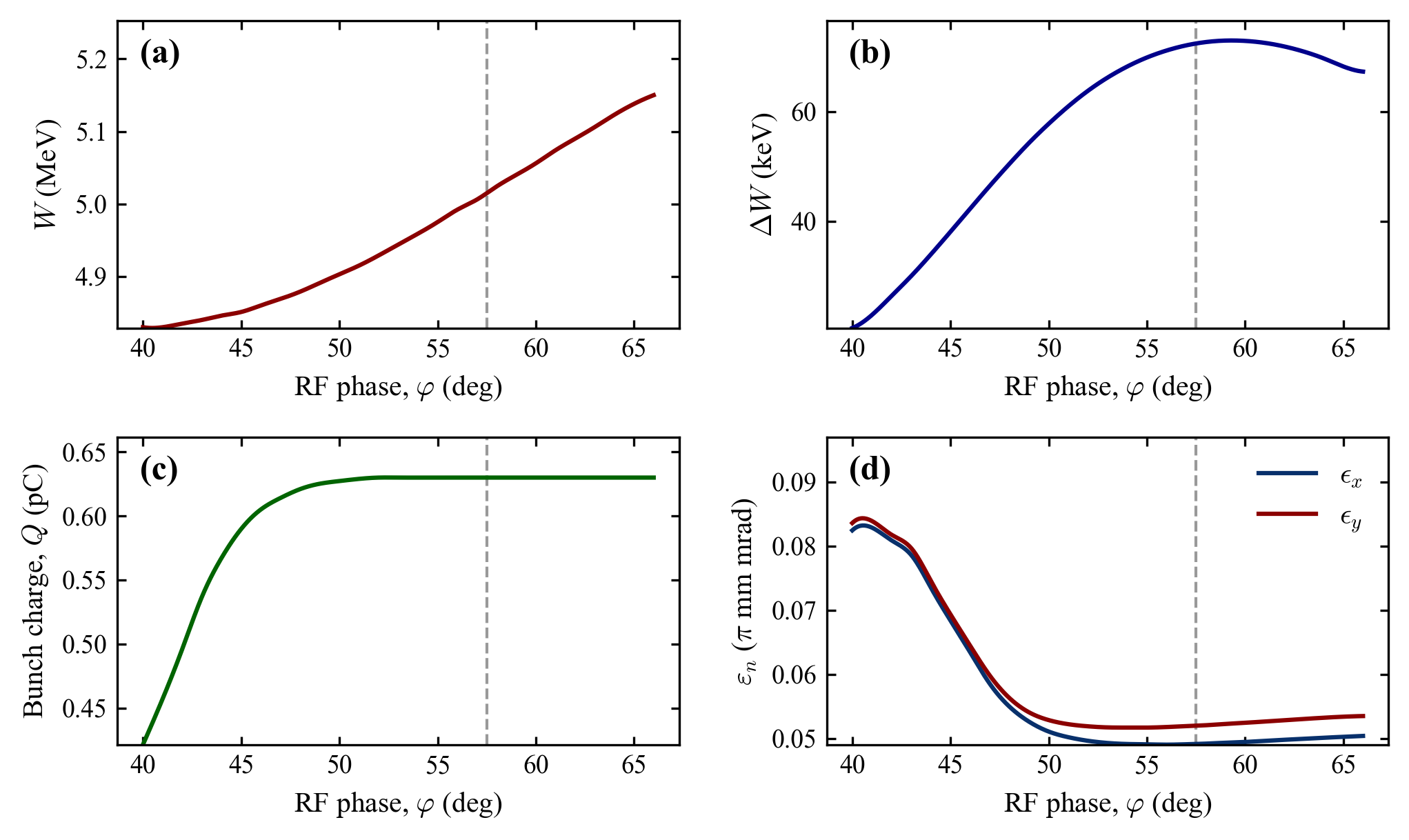}
        \caption{RF phase scan results obtained from beam dynamics simulations. The subplots display the dependence of the 
        (a) mean kinetic energy $W$, 
        (b) rms energy spread $\Delta W$, 
        (c) bunch charge, and 
        (d) transverse normalized emittances $\varepsilon_{x,y}$ on the operating RF phase $\varphi$. The vertical dashed gray line across all panels indicates the selected optimal 
        phase of $57.5^\circ$.}
        \label{fig:phase_scan}
    \end{figure*}

    The optimal RF phase was initially determined via simulations in CST Microwave Studio and subsequently matched to the corresponding value in ASTRA. When determining this optimal 
    phase, it is crucial to strike a balance between maximizing the beam energy gain, minimizing the energy spread, preserving the maximum transmitted charge, and achieving the 
    minimum transverse emittance. The detailed phase scan illustrating these dependencies at the exit of the RF gun is presented in Fig.~\ref{fig:phase_scan}(a--d). This evaluation 
    point was chosen because, for our parameters, the final emittance at the end of the entire beamline is expected to be governed primarily by the intrinsic thermal emittance at the 
    cathode. The optimal phase for our operating conditions is selected to be $57.5^\circ$. Although this working point does not yield the absolute maximum output energy or the 
    minimum energy spread, it lies at the center of an optimal phase window that ensures maximum charge transmission while maintaining the lowest possible beam emittance. If the 
    phase is increased to $60^\circ$--$65^\circ$, the final emittance in the beamline begins to degrade significantly---a predictable process whose onset is already illustrated in 
    Fig.~\ref{fig:phase_scan}(d)---while the output energy spread grows to approximately 70~keV. A similar overall degradation in beam quality is observed when the phase is decreased 
    by a comparable margin. The parameters used in subsequent simulations are summarized 
    in Table~\ref{tab:init_params}.

    \begin{figure*}[!ht]
        \centering
        \includegraphics[width=1\textwidth]{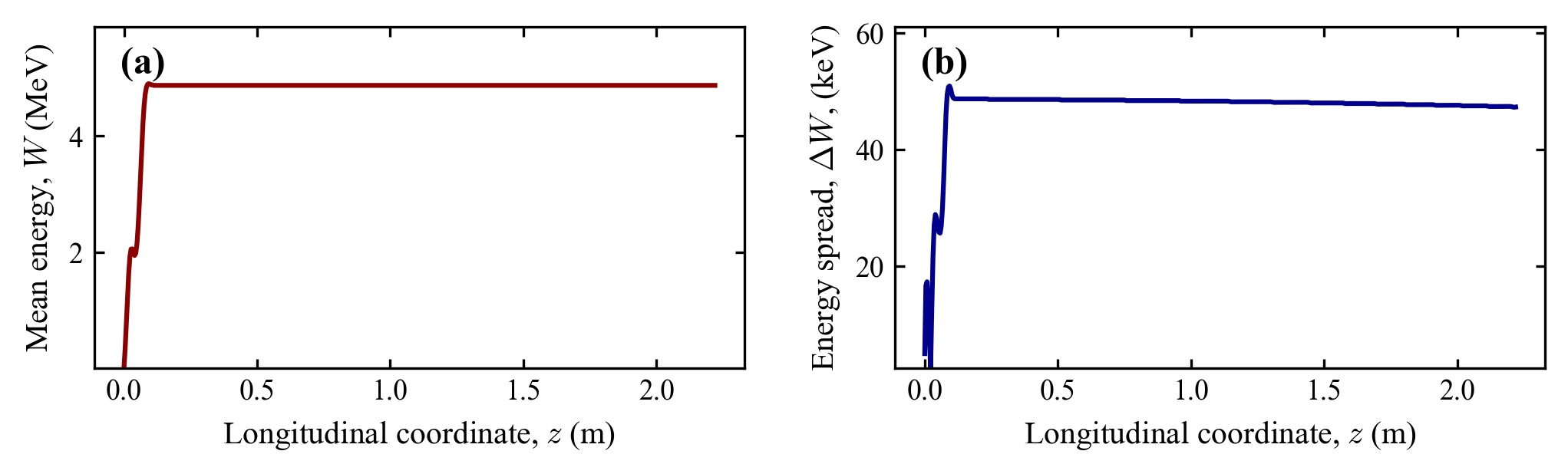}
        \caption{Evolution of the beam energy characteristics along the test bench beamline. 
        (a) Mean kinetic energy $W$ of the bunch. The beam exits the RF gun at $4.9$~MeV. 
        (b) RMS energy spread $\Delta W(z)$ during acceleration and subsequent drift, stabilizing at an average value of $47$~keV.}
        \label{fig:energy_gain}
    \end{figure*}

    \begin{figure}[!ht]
        \centering
        \includegraphics[width=0.5\textwidth]{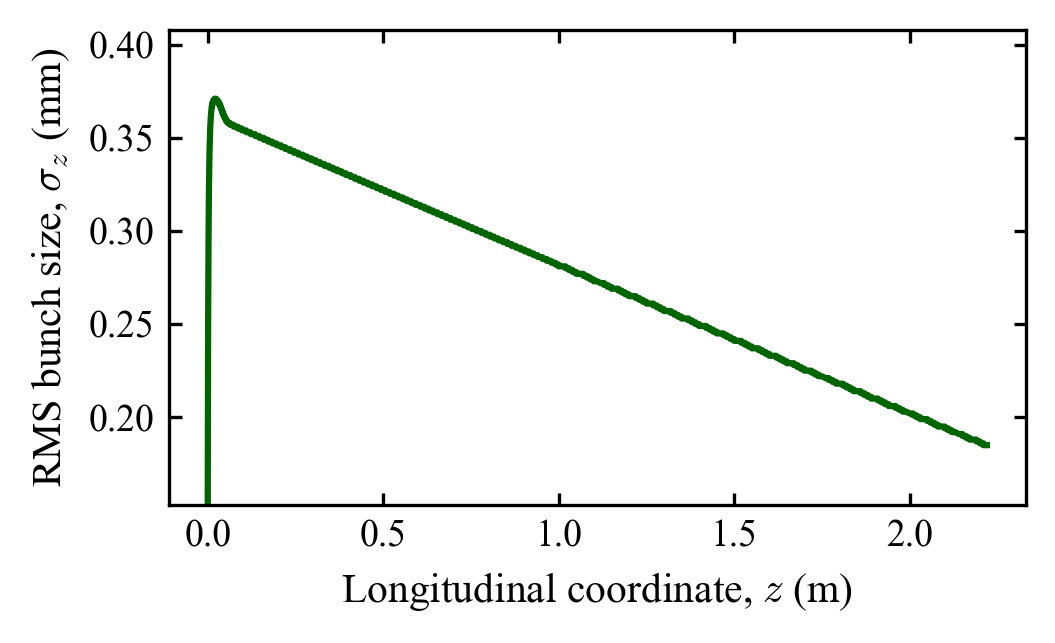}
        \caption{Evolution of the longitudinal rms bunch length $\sigma_z(z)$ along the test bench beamline. The linear decrease in the drift section indicates ballistic compression 
        induced by the RF field, with the bunch compressing to approximately $0.18$~mm at the exit.}
        \label{fig:sigma_rms_long}
    \end{figure}

    \begin{table}[!ht]
    \centering
    \caption{\label{tab:init_params}Initial parameters for beam dynamics simulations.}
    \begin{tabular}{lc}
        \toprule
        Parameter & Value \\
        \midrule
        Photon energy                           & 4.74 eV \\
        Work function (Cu)                      & 4.3 eV \\
        Mean kinetic energy                     & 0.29 eV \\
        RMS energy spread                       & 0.10 eV \\
        Pulse duration (RMS)                    & $4.225\ \mathrm{ps}$ \\
        RMS beam size at cathode                & $79\,\mathrm{\mu m}$ \\
        Normalized emittance                    & $46\ \pi \cdot \mu\mathrm{m\cdot mrad}$ \\
        Number of emitted electrons             & $3.95 \times 10^6$ \\
        Bunch charge                            & 0.63 pC \\
        Optimal RF phase                        & $57.5^\circ$ \\
        ASTRA tracking length (to dipole)       & $2.222\ \mathrm{m}$ \\
        Total beamline length                   & $2.9\ \mathrm{m}$ \\
        \bottomrule
    \end{tabular}
    \end{table}

    Fig.~\ref{fig:energy_gain} presents the simulated mean kinetic energy along the beamline. The characteristic two-peak structure reflects the axial field distribution of the 
    1.5-cell S-band RF gun. The beam exits the gun with $W=4.9 $~MeV after which the energy remains constant in the drift.

    The longitudinal rms bunch length $\sigma_z$ (Fig.~\ref{fig:sigma_rms_long}) exhibits a rapid initial increase immediately after emission. However, downstream of the RF cavity, 
    the bunch undergoes ballistic compression rather than expansion. At the selected optimal RF phase of $57.5^\circ$, the accelerating field imparts a correlated energy spread 
    (longitudinal chirp) to the electron bunch, causing the tail of the bunch to acquire a slightly higher velocity than its head. As a result, the bunch length decreases linearly as 
    it travels along the drift section of the test bench, compressing to $\sigma_z \approx 0.18$~mm at the beamline exit.
    
    \begin{figure*}[!ht]
        \centering
        \includegraphics[width=1\textwidth]{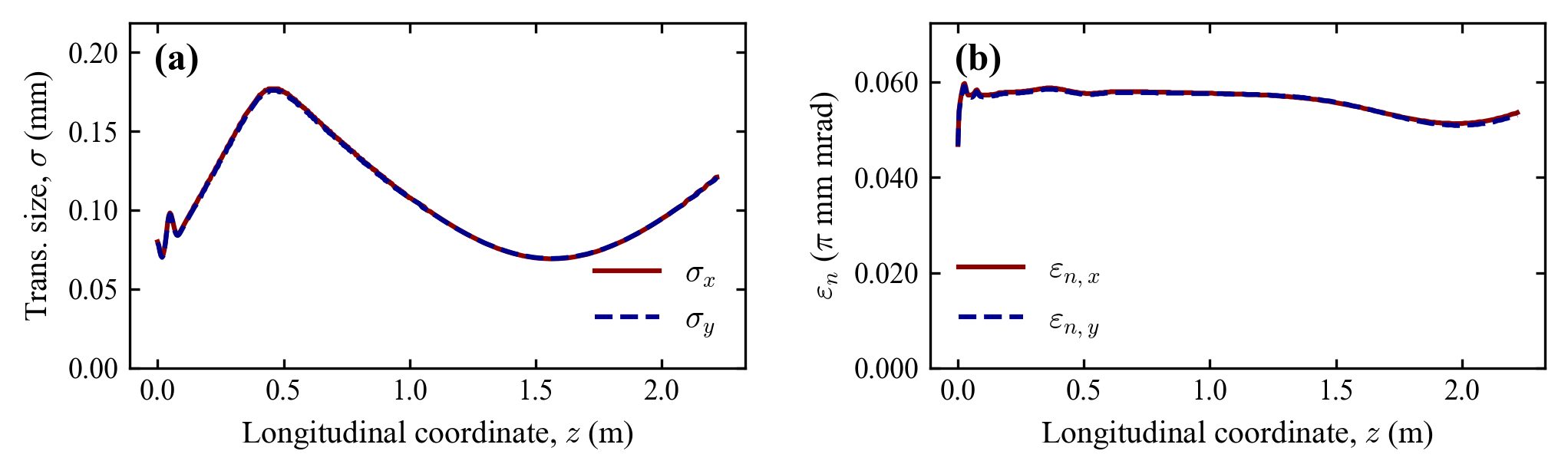}
        \caption{Evolution of the transverse beam characteristics along the test bench beamline. 
        (a) Transverse RMS beam size $\sigma_{x,y}(z)$. The beam size is $69~\mu$m at the target area and $108~\mu$m at the beamline exit. 
        (b) Normalized transverse emittance $\varepsilon_{n,x,y}(z)$, reaching $56\ \pi\cdot\mathrm{\mu m\cdot mrad}$ at the end of the beamline.}
        \label{fig:transverse_dynamics}
    \end{figure*}

    A compensation solenoid is installed immediately downstream of the photocathode to control the rapid transverse expansion of the low-energy beam. As noted earlier, without this 
    external magnetic focusing, severe particle losses occur within the first 25~cm of the beamline due to the restrictive geometry of the coaxial input structure. The measured field 
    map (Fig.~\ref{fig:solenoid_field}) was incorporated into the ASTRA environment. During the beam dynamics simulations, the field amplitude was scaled and optimized
    iteratively to match the beam envelope to the subsequent drift section. The resulting optimized simulation operating point corresponds to a peak magnetic field of
    $B_z = 174$~mT at $z = 85$~mm from the cathode plane. This specific field configuration not only ensures complete 
    charge transmission through the RF-gun apertures but also establishes the necessary phase-space correlations for proper emittance compensation and stable downstream transport.
    
    Figure~\ref{fig:transverse_dynamics}(a) shows the rms transverse beam size along the beamline. After a short adjustment region immediately downstream of the gun, 
    the beam forms a waist of $\sigma_{x,y} = 69\ \mu\mathrm{m}$ at the target area. The formation of this local minimum clearly demonstrates the effectiveness of the focusing 
    solenoid, which successfully minimizes the beam spot exactly where the experimental target will be located. Beyond this region, the beam gradually expands to $108\ \mu\mathrm{m}$ 
    at the beamline exit. The symmetry between $x$ and $y$ confirms the predominantly axisymmetric dynamics.

    The normalized transverse emittance evolution is shown in Fig.~\ref{fig:transverse_dynamics}(b). Inside the solenoid region, the emittance exhibits a pronounced increase, 
    reaching approximately $\varepsilon_{n, x, y} \approx 56\ \pi\cdot\mathrm{\mu m\cdot mrad}$, which reflects the magnetization acquired by the beam in the presence of a 
    non-zero axial magnetic field at the cathode. The solenoidal field introduces transverse correlations associated with angular momentum, temporarily increasing the projected 
    emittance. The interplay between the solenoidal field and the transverse RF fields of the 1.5-cell $\pi$-mode gun modifies the natural cancellation of RF-induced 
    correlations that occurs in the zero-solenoid case~\cite{Floettmann15}.

    It is also necessary to address the macroscopic magnetization introduced by the cathode solenoid in the context of the generalized Busch theorem. While the magnetic field 
    imparts a canonical angular momentum to the macroscopic beam envelope, the generation of a significant \textit{intrinsic} (quantized) OAM projection solely via magnetic 
    transition typically requires extreme field strengths on the order of $1$~T or higher~\cite{Karlovets21}. Because our peak solenoidal field is relatively modest ($174$~mT), 
    any magnetically induced intrinsic OAM projection carries negligible weight compared to the large topological charge (up to $\ell = 64$) imprinted by the structured UV laser. 
    Furthermore, downstream of the magnetic elements, the kinetic and canonical orbital angular momenta coincide, ensuring that the topological charge $\ell$ is a well-defined, 
    gauge-invariant observable. As discussed later, this integer topological charge can be experimentally verified via diffraction on a triangular aperture~\cite{Maksimov25}.

\begin{figure}[!ht]
        \centering
        \includegraphics[width=0.75\textwidth]{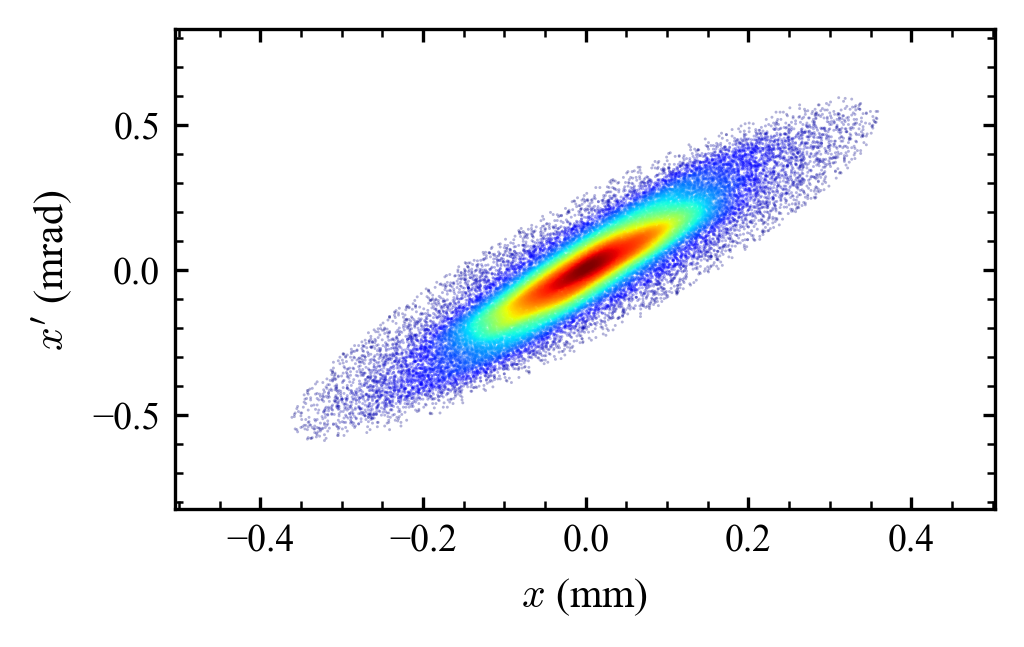}
        \caption{Simulated transverse phase space of the electron beam at the end of the beamline.
        The $x$–$x'$ projection exhibits an approximately ideal phase-space ellipse, indicating stable beam formation.}
        \label{fig:trans_phase_space}
    \end{figure}

    Figure~\ref{fig:trans_phase_space} shows the transverse phase-space distribution at the end of the beamline. The $x$–$x'$ projection forms an approximately elliptical distribution 
    with no pronounced filamentation, consistent with controlled envelope evolution. Due to axisymmetry, an identical structure is expected in the $y$–$y'$ plane. The overall 
    picture confirms that, at the present low charge, the beam dynamics is dominated by RF-induced correlations and controlled solenoidal magnetization, while space-charge 
    effects remain subdominant after the initial acceleration stage.
    
    Operating at a bunch charge of $0.63\ \mathrm{pC}$ places the test bench in a regime where thermal emittance remains a dominant contribution. Furthermore, the 
    total charge of an 80-bunch train ($\approx 50.4\ \mathrm{pC}$) induces negligible cavity beam loading ($<0.01\%$ voltage drop), ensuring nearly identical beam dynamics for all 
    bunches. The simulated normalized rms emittance at the end of the beamline is $\varepsilon_{n, x,y} \approx 56\ \pi\cdot\mathrm{\mu m\cdot mrad}$. 
    This result indicates that the final emittance remains close to the intrinsic thermal-emittance scale of the photoemission source, with no substantial additional degradation from 
    RF-induced correlations or space-charge effects.

\subsection{\label{sec:dipole_transport}Dipole transport and Twiss parameter evolution}

    To accurately simulate the subsequent beam transport through the sector dipole magnet, the particle distribution at $z = 2.222$~m (the entrance of the dipole 
    chamber) was transferred to OCELOT~\cite{Agapov14}. This transition is necessitated by the inherent limitations of the ASTRA code. Specifically, ASTRA relies strictly on a global 
    Cartesian coordinate system and lacks the capability to analytically separate the geometric beam envelope (Twiss $\beta$-function) from the chromatic broadening 
    (dispersion function $D$). OCELOT, conversely, utilizes a curvilinear coordinate system and matrix formalism, enabling a rigorous analytical description of the Twiss parameter 
    evolution along the reference trajectory.

\begin{figure}[!ht]
        \centering
        \includegraphics[width=0.5\textwidth]{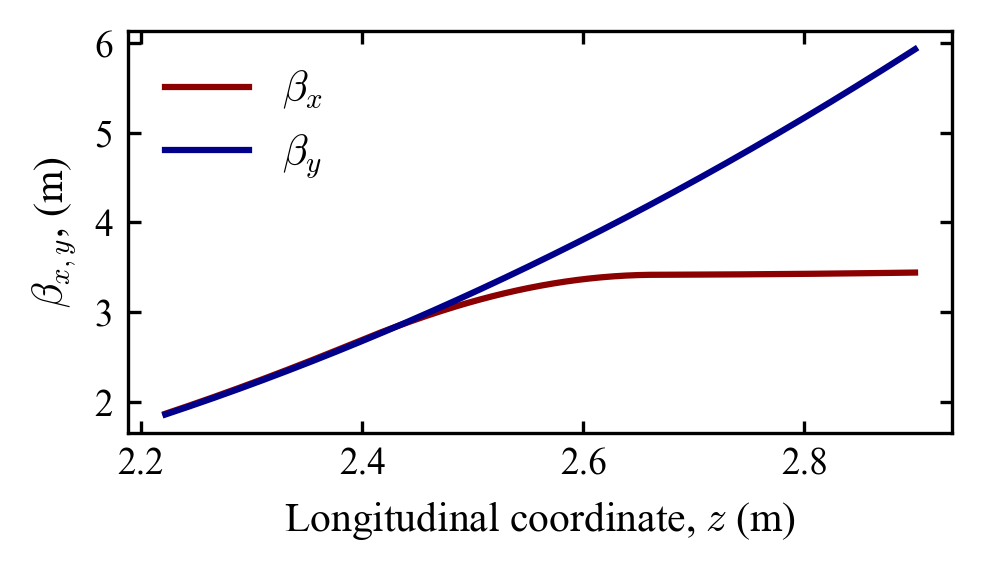}
        \caption{Evolution of the Twiss $\beta$-functions along the reference trajectory in the dipole section. The horizontal weak focusing of the sector dipole suppresses the 
        growth of $\beta_x$, while $\beta_y$ continues to evolve identically to a drift space.}
        \label{fig:dipole_beta}
    \end{figure}

    The evolution of the Twiss $\beta$-functions through the dipole region is shown in Fig.~\ref{fig:dipole_beta}. Initially, both $\beta_x$ and $\beta_y$ grow symmetrically as the 
    beam propagates through the drift space before the magnet. Upon entering the sector dipole, the beam experiences horizontal weak focusing. Consequently, the growth of the 
    horizontal envelope function $\beta_x$ is suppressed, whereas the vertical function $\beta_y$ continues to diverge since a pure sector magnet provides no vertical 
    focusing.

    \begin{figure*}[!ht]
        \centering
        \includegraphics[width=1\textwidth]{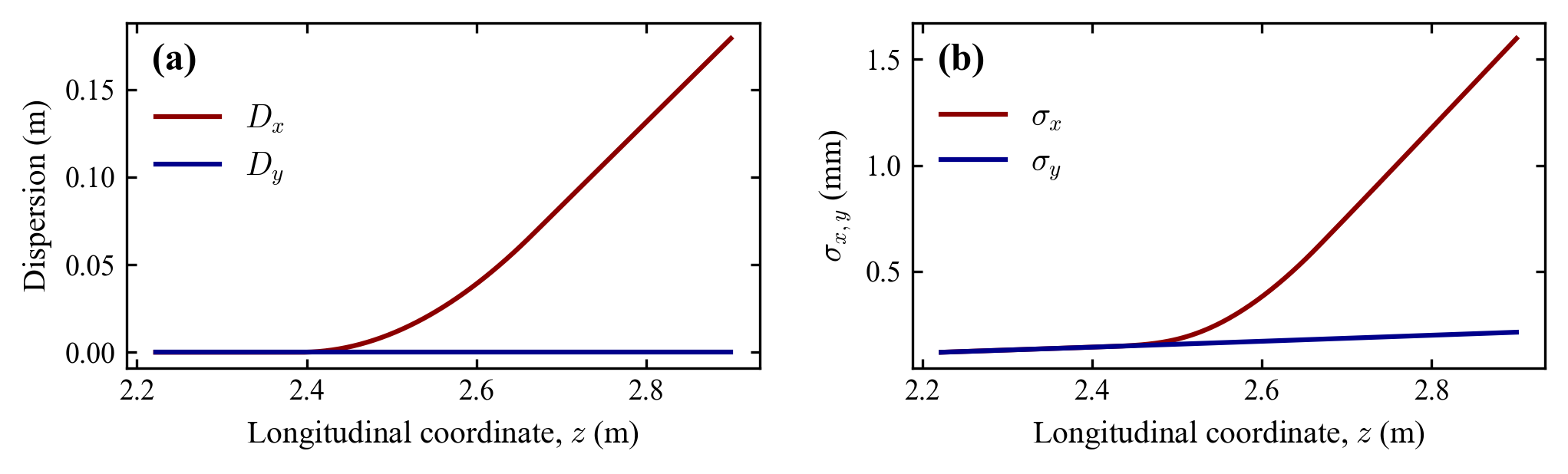}
        \caption{Evolution of the dispersive and transverse beam parameters along the sector dipole section. 
        (a) Generated horizontal ($D_x$) and vertical ($D_y$) dispersion functions. 
        (b) Total transverse rms beam spot sizes $\sigma_{x,y}$. The rapid growth of the horizontal spot size $\sigma_x$ is entirely dominated by chromatic broadening (dispersion), 
        superseding the geometric weak focusing effect.}
        \label{fig:dipole_dispersion_sigma}
    \end{figure*}

    Figure~\ref{fig:dipole_dispersion_sigma} illustrates the evolution of the dispersion functions $D_{x,y}$ [Fig.~\ref{fig:dipole_dispersion_sigma}(a)] and the 
    total rms beam spot sizes $\sigma_{x,y}$ [Fig.~\ref{fig:dipole_dispersion_sigma}(b)] inside the dipole chamber. As the beam is deflected, the horizontal dispersion $D_x$ increases 
    significantly, while $D_y$ remains zero. It is important to note that $\sigma_x$ in this context represents the total observable spot size on a virtual diagnostic screen, which is 
    defined by the convolution of the geometric beam size and the chromatic broadening~\cite{Wiedemann15}:

    \begin{equation}
        \sigma_x = \sqrt{\varepsilon_x \beta_x + (D_x \Delta W)^2},
    \end{equation}

    Despite the geometric focusing in the horizontal plane indicated by the $\beta_x$ plateau, the total 
    horizontal spot size $\sigma_x$ undergoes rapid growth. This expansion is completely dominated by the dispersive term due to the energy spread acquired during the RF acceleration. 
    Because this chromatic broadening significantly exceeds the purely geometric contribution ($D_x \Delta W \gg \sqrt{\varepsilon_x \beta_x}$), the horizontal spatial profile on 
    the diagnostic screen acts as a direct macroscopic map of the bunch energy distribution. Consequently, this dispersive mapping is exactly what enables the sector dipole to 
    function as a spectrometer, allowing for the precise experimental reconstruction of the longitudinal energy spectrum based on the measured transverse spot size.
    
\section{Quantum beam dynamics}\label{sec:quantum}

    The classical beam dynamics simulations presented above demonstrate that the photoinjector can operate in a regime of low bunch charge and correspondingly weak 
    space-charge forces, yielding a stable, low-emittance beam with minimal RF-induced phase-space distortions. This defines the classical operating regime used below. 
    In this regime, it becomes essential to evaluate whether structured quantum states -- specifically, vortex electrons -- can be accelerated while retaining their characteristic 
    transverse vortex structure under realistic beam-dynamics conditions. 
    Motivated by the experimentally accessible values of orbital angular momentum that can be transferred by our UV vortex optics, we now turn to the quantum description of 
    single-electron Laguerre--Gaussian states in the RF fields of the photoinjector.

    It is important to clarify the validity limits of the single-electron quantum framework in our context. A bunch charge of $Q = 0.63$~pC corresponds to 
    approximately $4 \times 10^6$ electrons. While the classical ASTRA simulations demonstrate that gross space-charge emittance growth is strongly mitigated in this 
    ultralow-charge regime, stochastic Coulomb interactions will inevitably contribute to beam decoherence. Currently, a rigorous many-body quantum-kinetic treatment of 
    collective Coulomb effects for structured macroscopic wave packets is beyond existing theoretical models. Therefore, the single-particle quantum dynamics presented here 
    should be interpreted as an idealized baseline -- a theoretical upper bound on single-electron spatial-structure preservation;
    it does not establish phase coherence or OAM preservation. Operating the photoinjector at this low charge is 
    the closest experimentally accessible approximation to a weakly collective regime for the purpose of this baseline estimate, serving as a necessary prerequisite for observing 
    these delicate quantum states.

\subsection{\label{sec:vortex_electrons}Vortex electrons}

    In general, a vortex electron beam is a free-electron state whose wave function can be represented in terms of an LG mode, an exact non-stationary solution of the Schrödinger 
    equation and a close analogue of optical vortex modes~\cite{Allen92}. Its transverse wave function can be written as~\cite{Karlovets21}
    
    \begin{equation}
        \begin{split}
            \psi_{\ell, n}(\boldsymbol{\rho}, t) &= N \frac{\rho^{\vert \ell \vert}}{(\sigma_{\perp}(t))^{\vert \ell \vert +1}} 
            L_n^{\vert \ell \vert} \left(\frac{\rho^2}{(\sigma_{\perp}(t))^2} \right) \\
            &\quad \times \exp\Biggl[i \ell \phi_r - i(2n + \vert \ell \vert +1) \arctan\left(\frac{t}{t_d}\right) \\
            &\qquad - \frac{\rho^2}{2(\sigma_{\perp}(t))^2} \left(1 - i \frac{t}{t_d} \right) \Biggr]
        \end{split}
        \end{equation}
    where the normalization constant $N$ is chosen such that
    
    \begin{equation}
        \int d^2 x\, \vert \psi_{\ell,n}(\boldsymbol{\rho},t) \vert^2 = 1.
    \end{equation}
    Here, $\ell = 0, \pm 1, \pm 2,...,$  is the OAM in units of $\hbar$, $n = 0, 1, 2,...,$ is the radial quantum number, $L_n^{|\ell|}$ denotes the generalized 
    Laguerre polynomial, $\sigma_{\perp}(t)$ is the Gaussian-envelope width and equals the transverse rms radius for the fundamental mode $n=\ell=0$, and $t_d$ is the 
    characteristic diffraction time. The azimuthal phase factor $\exp(i\ell\phi_r)$ makes the state an eigenfunction of the OAM operator,
    \[
    \hat L_z\,\psi_{\ell,n}= -i \hbar \frac{\partial}{\partial \phi_r} \psi_{\ell,n}=\ell\hbar\,\psi_{\ell,n},
    \]
    and it is the defining feature of vortex beams: it produces the characteristic helical phase front and ensures that each electron carries an orbital angular momentum projection $\ell\hbar$ with respect to the propagation axis.

    One mechanism for generating vortex electrons in an RF photoinjector at an accelerator facility is laser-driven emission using structured light.
    Theory shows that an atom placed on the axis of an optical vortex transfers a well-defined OAM projection to the emitted electron, and this behaviour persists for ensembles~\cite{Pavlov24, Kazinski25}. 
    Alternative schemes based on electron diffraction~\cite{Uchida10, Verbeeck10, Blackburn14} are impractical in such setups.

\subsection{\label{sec:quantum_evolution}Quantum evolution of a single-electron wave packet in the accelerating field}

    To analyze the generation of relativistic vortex electrons carrying OAM, we follow the quantum evolution of a single-electron wave packet during acceleration. 
    As a convenient observable we consider the mean-squared transverse radius $\langle \rho^2 \rangle(\langle z \rangle )$, which for an LG wave packet is related to the transverse 
    rms width via~\cite{Karlovets21}
    \begin{equation}
        \langle \rho^2 \rangle (\langle z \rangle ) = \sigma_\perp^2(\langle z \rangle )\,(2n + \vert \ell \vert +1).
    \end{equation}
    Throughout this section, the angular brackets denote quantum expectation values over the single-electron state. In the main text we focus on two representative cases:
    \begin{enumerate}
    \item free-space propagation over the RF-photogun length;
    \item acceleration in the realistic standing-wave RF field of the photoinjector.
    \end{enumerate}
     
    For reference, the nonrelativistic law of transverse quantum spreading for free wave packets~\cite{Karlovets21} can be written either as a function of the mean 
    propagation distance $\langle z\rangle$ or of the corresponding time of flight $t(\langle z\rangle)$:
    \begin{equation}
        \label{eq:general_law}
        \langle \rho^2 \rangle(\langle z \rangle)
        = \langle \rho^2 \rangle(0) \left(1 + \frac{\langle z \rangle^2}{z_R^2}\right)
        = \langle \rho^2 \rangle(0) \left(1 + \frac{t^2(\langle z \rangle)}{t_d^2}\right).
    \end{equation}
    Here $z_R$ is the Rayleigh length and $t_d$ is the diffraction time~\cite{Karlovets21},
    \begin{equation}
        \label{eq:rayleigh}
        z_R = \frac{2 \pi}{M} \frac{\langle \rho^2 \rangle(0)}{\lambda_{\mathrm{dB}}}, \qquad
        t_d = \frac{m\,\langle \rho^{2}\rangle(0)}{\hbar M},
    \end{equation}
    where $M = 2n + |\ell| + 1$ is the beam-quality factor of the packet and \(\lambda_{\mathrm{dB}}=2\pi\hbar/p_z\) is the longitudinal de~Broglie wavelength.
    Within the uncorrelated LG family, Eqs.~\eqref{eq:general_law} and \eqref{eq:rayleigh} depend on $(n,\ell)$ through $M$ and the initial second moment
    $\langle \rho^{2}\rangle(0)$. The initial rms radius fully specifies the spreading within this family, whereas a general coherent packet also requires the initial transverse-momentum variance and position--momentum covariance.

    In the far-field regime, $\langle z \rangle \gg z_R$, Eq.~\eqref{eq:general_law} reduces to a simple relation, often referred to as a van Cittert–Zernike–type 
    theorem~\cite{Karlovets21}, between the measured packet width at the detector and the initial width near the source,
    \begin{equation}
    \label{eq:far_field}
        \sqrt{\langle \rho^2 \rangle (0)}
        = \frac{\langle z \rangle\, \lambda_{\mathrm{dB}}}{2 \pi\, \sqrt{\langle \rho^2 \rangle (\langle z \rangle)}}\, M.
    \end{equation}

    For $z_R/\langle z\rangle\ll1$, the first correction to the far-field result is
    \begin{equation}
        \begin{split}
            \sqrt{\langle \rho^2 \rangle (\langle z \rangle)}
            \simeq {} & \frac{\langle z \rangle\, \lambda_{\mathrm{dB}}}{2 \pi\, \sqrt{\langle \rho^2 \rangle (0)}}\, M \\
            & \times \left[1 + \frac{1}{2}
            \left(
                \frac{2 \pi\, \langle \rho^2 \rangle (0)}{M\langle z \rangle\, \lambda_{\mathrm{dB}}}
            \right)^{2}\right].
        \end{split}
    \end{equation}

    For an initial rms transverse radius $\sqrt{\langle \rho^2 \rangle (0)} = 1\ \mathrm{nm}$, typical for an electron packet near the emission region~\cite{Karlovets21, Ehberger15}, 
    with $n = 3$ and $\ell = 1$ (OAM $=\ell\hbar$) and a drift length of $30\ \mathrm{cm}$, Eq.~\eqref{eq:general_law} predicts free-space spreading to an rms transverse width of 
    about $0.8\ \mathrm{m}$ for an initial longitudinal-carrier kinetic energy of $0.3\ \mathrm{eV}$. Next, we include acceleration.

\paragraph*{RF standing-wave acceleration.}
    In the S-band photoinjector electrons are accelerated in a standing-wave RF field.
    Along the axis, where the packet propagates, the longitudinal electric field can be written as~\cite{Wiedemann15, Chao20, Lee21}
    \begin{equation}
    E_z(z) = E_{z0}\,\sin\!\left(\omega \frac{z}{v} + \phi\right),
    \label{eq:Ez_RF_onaxis}
    \end{equation}
    where $E_{z0}$ is the peak amplitude, $\omega$ is the RF angular frequency, and $\phi$ is the injection phase. The same sine convention is used throughout.
    In the constant-velocity gap approximation, the corresponding kinetic-energy gain over an accelerating gap of length $d$ is
    \begin{equation}
        \Delta E_{\rm kin}
        = q E_{z0}
        \int_{-d/2}^{d/2}\!
        \sin\!\left(\omega\frac{z}{v}+\phi\right)\,dz,
        \label{eq:DeltaE_RF}
    \end{equation}
    where $q=-e$ and $e>0$. This expression shows the phase dependence in the constant-velocity approximation. The numerical calculation uses the simulated on-axis energy $W(z)$ and the RF phase convention implemented in ASTRA.
    
  \paragraph*{Relativistic longitudinal-clock model.}
    In the TM$_{010}$ standing-wave mode the on-axis electromagnetic field is predominantly
    longitudinal.  Near the axis the field admits the expansions
    \begin{align}
        E_z(\rho,z,t) &= E_z(0,z,t)
        + \mathcal{O}\!\left[E_{z0}\left(\frac{\rho}{R_{\mathrm{cav}}}\right)^2\right],\\
        B_\varphi(\rho,z,t) &=
        \mathcal{O}\!\left[\frac{E_{z0}}{c}\frac{\rho}{R_{\mathrm{cav}}}\right],
    \end{align}
    where $R_{\mathrm{cav}}$ is the cavity radius. At the emission plane, the initial rms radius
    $\sqrt{\langle\rho^2\rangle(0)}=1$~nm gives $\sqrt{\langle\rho^2\rangle(0)}/R_{\mathrm{cav}}\lesssim10^{-8}$. Consequently, the initial transverse magnetic force relative to the peak longitudinal electric force satisfies $|qv_zB_\varphi|/|qE_{z0}|\ll10^{-7}$, so the transverse magnetic force is negligible at the source.
    For the maps below, the on-axis RF solution supplies the longitudinal energy $W(z)$. Transverse RF and solenoid forces on the internal LG packet lie beyond the present longitudinal approximation. For a transverse momentum $p_\perp$, the positive-energy dispersion is
    \begin{equation}
        \varepsilon(p_\perp,z)
        =mc^2\sqrt{\gamma^2(z)+\left(\frac{p_\perp}{mc}\right)^2}.
        \label{eq:positive_energy_dispersion}
    \end{equation}
    \noindent\textit{Longitudinal evolution and time-of-flight.}
    The kinetic energy, Lorentz factor, and normalized longitudinal velocity of the carrier are
    \begin{equation}
        \begin{aligned}
            W(z)&=W_0+\Delta E_{\rm kin}(z),
            &\gamma(z)&=1+\frac{W(z)}{mc^2},\\
            \beta(z)&=\sqrt{1-\frac{1}{\gamma^2(z)}}.
        \end{aligned}
    \end{equation}

    Here $W_0$ is the initial longitudinal kinetic energy, and $\Delta E_{\rm kin}(z)$ is the cumulative energy gain obtained from the simulated on-axis energy evolution.

    The flight time to a plane at $z$ follows from
    \begin{equation}
        t_{\rm lab}(z) = \int_0^{z}\frac{dz'}{\beta(z')\,c}.
        \label{eq:tof_RF}
    \end{equation}

    The laboratory time $t_{\rm lab}(z)$ is retained for the RF phase. The transverse evolution is calculated from the positive-energy dispersion in Eq.~\eqref{eq:positive_energy_dispersion} using the simulated longitudinal energy $W(z)$. In the paraxial limit, $p_\perp/p_z\ll1$, this procedure reduces to the longitudinal diffraction clock obtained for stationary fields by Meng \textit{et al.}~\cite{z6j7-grs2} and by Filina and Baturin~\cite{FilinaBaturin2026} in the adiabatic/WKB regime.

\begin{figure*}[!ht]
    \centering

    \begin{subfigure}[t]{0.45\textwidth}
        \centering
        \includegraphics[width=\textwidth]{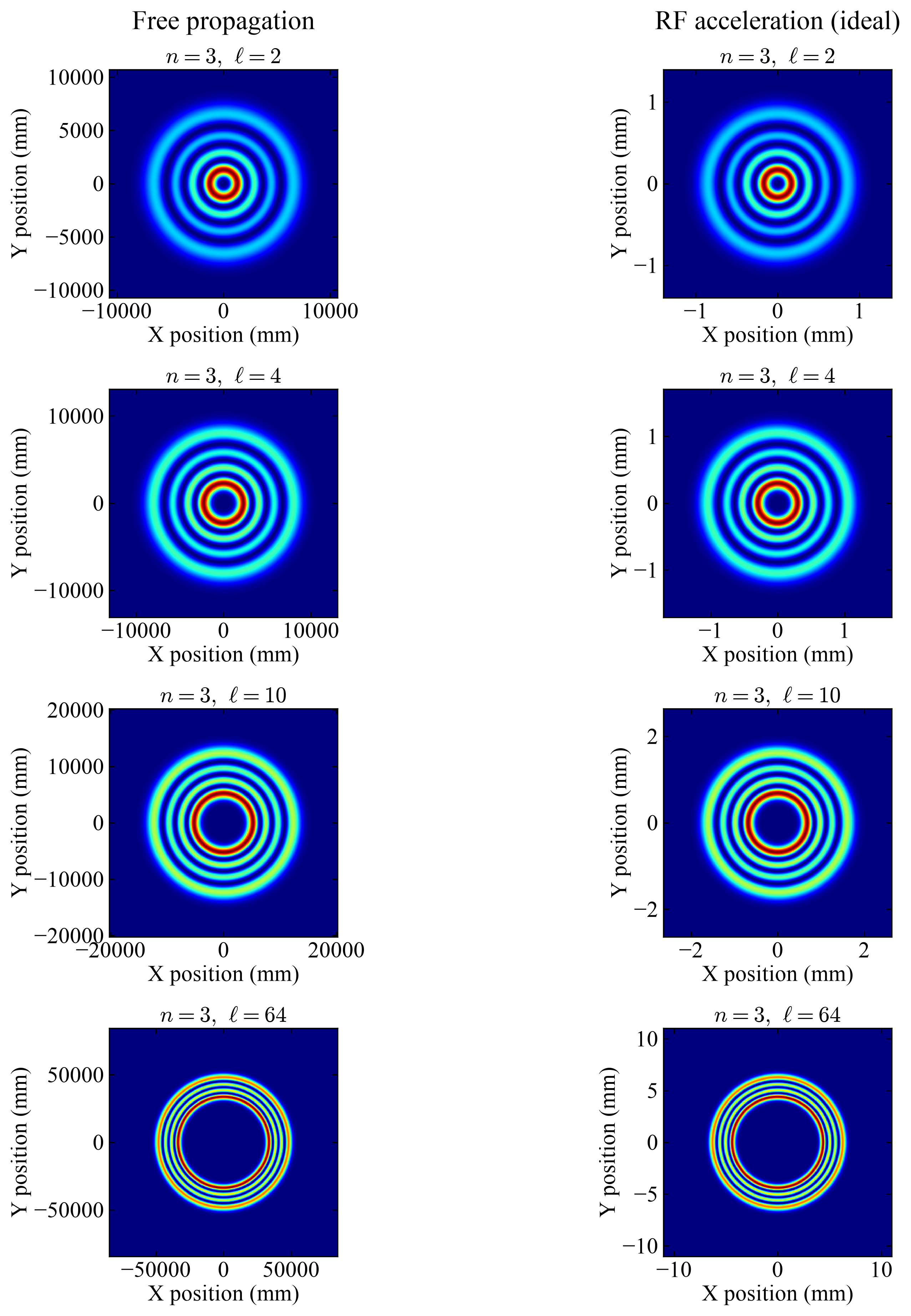}
        \caption{Dependence on the OAM value at the detector plane $z = 1.9$~m.}        
        \label{fig:rf_oam_scan}
    \end{subfigure}
    \hfill
    \begin{subfigure}[t]{0.45\textwidth}
        \centering
        \includegraphics[width=\textwidth]{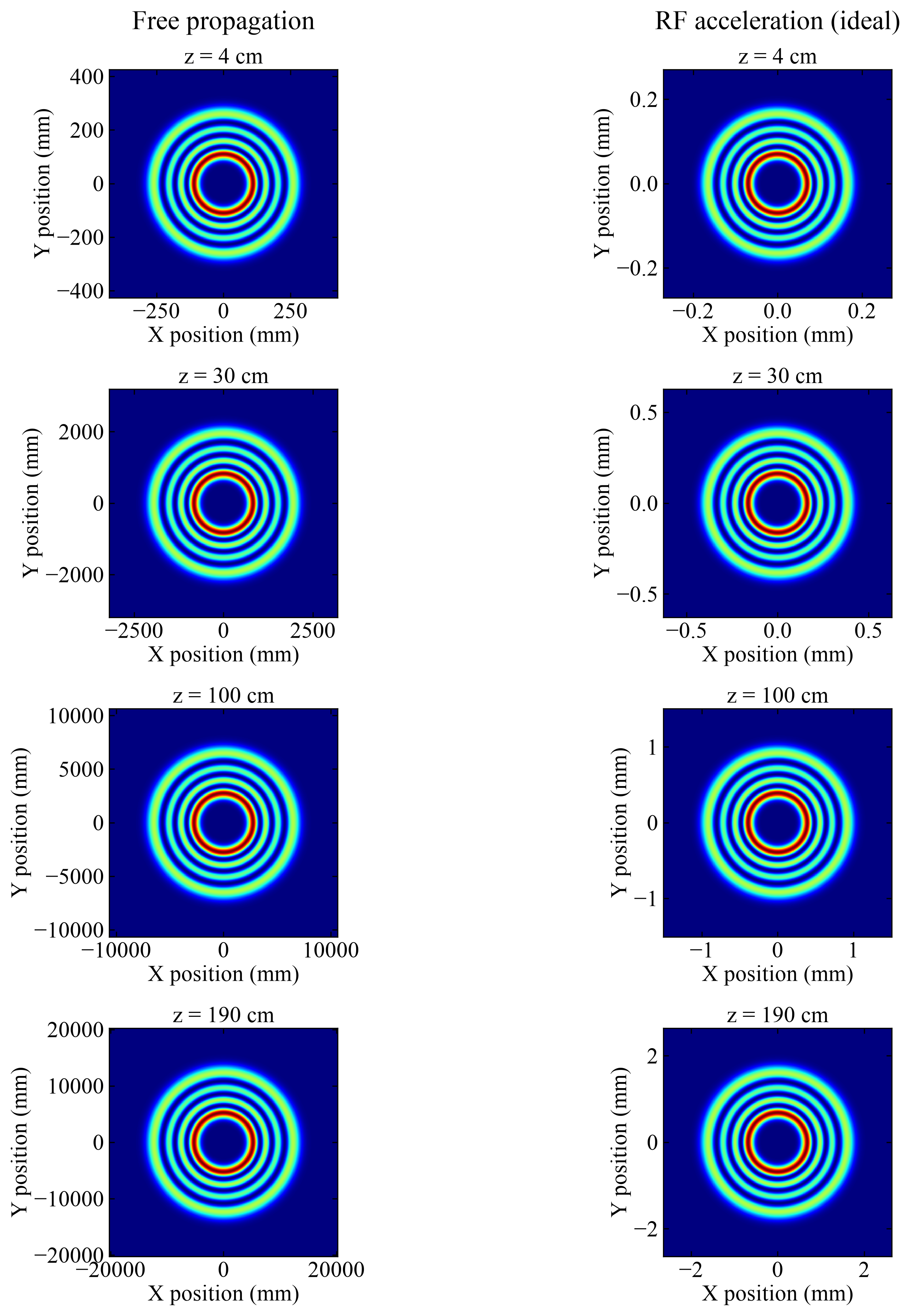}
        \caption{Dependence on the observation plane for the packet with $n=3$ and $\ell=10$.}
        \label{fig:rf_z_scan}
    \end{subfigure}

    \caption{Transverse probability-density maps $|\psi(x,y)|^2$ of Laguerre--Gaussian packets in free propagation (left column) and under ideal RF acceleration in the standing-wave field of the S-band 
    photogun (right column). In panel (a), the OAM dependence is shown for $n=3$ and $\ell=2,4,10,64$ at the detector plane $z=1.9$~m. In panel (b), the longitudinal evolution is 
    shown for the reference packet with $n=3$ and $\ell=10$ at $z=4$, $30$, $100$, and $190$~cm. In all panels, the initial rms transverse radius is 
    $\sqrt{\langle \rho^2\rangle(0)}=1$~nm and the initial longitudinal-carrier kinetic energy is $W_0=0.39$~eV. Each probability density is normalized to unit integral; the linear 
    color scale is adjusted independently for each panel. For the reference packet with $n=3$ and $\ell=10$, RF acceleration at the detector plane suppresses the free-space transverse 
    rms spreading by a factor of approximately $7.67\times10^3$ while preserving the characteristic annular structure.}

    \label{fig:rf_quantum}
    \end{figure*}

    For the photoinjector parameters used in this work, relativistic longitudinal motion strongly reduces the effective transverse spreading compared with free propagation. For the reference packet with $n=3$ and $\ell=10$, the free-space-to-RF suppression factors of the transverse rms size are $1.57\times10^3$, $5.09\times10^3$, $7.06\times10^3$, and $7.67\times10^3$ at $4$, $30$, $100$, and $190$~cm, respectively, as illustrated in Fig.~\ref{fig:rf_z_scan}. The ideal on-axis carrier reaches a kinetic energy of $4.9$~MeV at the detector plane $z=1.9$~m. The static-field analyses of Refs.~\cite{z6j7-grs2,FilinaBaturin2026} provide a qualitative check of this longitudinal model; full RF-plus-solenoid transport requires the corresponding transverse-field extension.

    To estimate nonideal broadening, we use the available nominal ASTRA centroid distributions as a conditional benchmark. The internal LG intensity is convolved with
    these distributions and averaged over RF amplitude and phase fluctuations with a relative rms amplitude jitter of 2\% and an rms phase jitter of $1^{\circ}$. The same nominal
    centroid distribution is used for every jitter realization, so the averaging varies the RF amplitude and phase while retaining the factorized internal-packet model. The ASTRA
    centroid distribution contains the classical solenoid response, whereas the internal LG packet is propagated with the longitudinal RF model.

\begin{figure*}[!ht]
    \centering

    \begin{subfigure}[t]{0.45\textwidth}
        \centering
        \includegraphics[width=\textwidth]{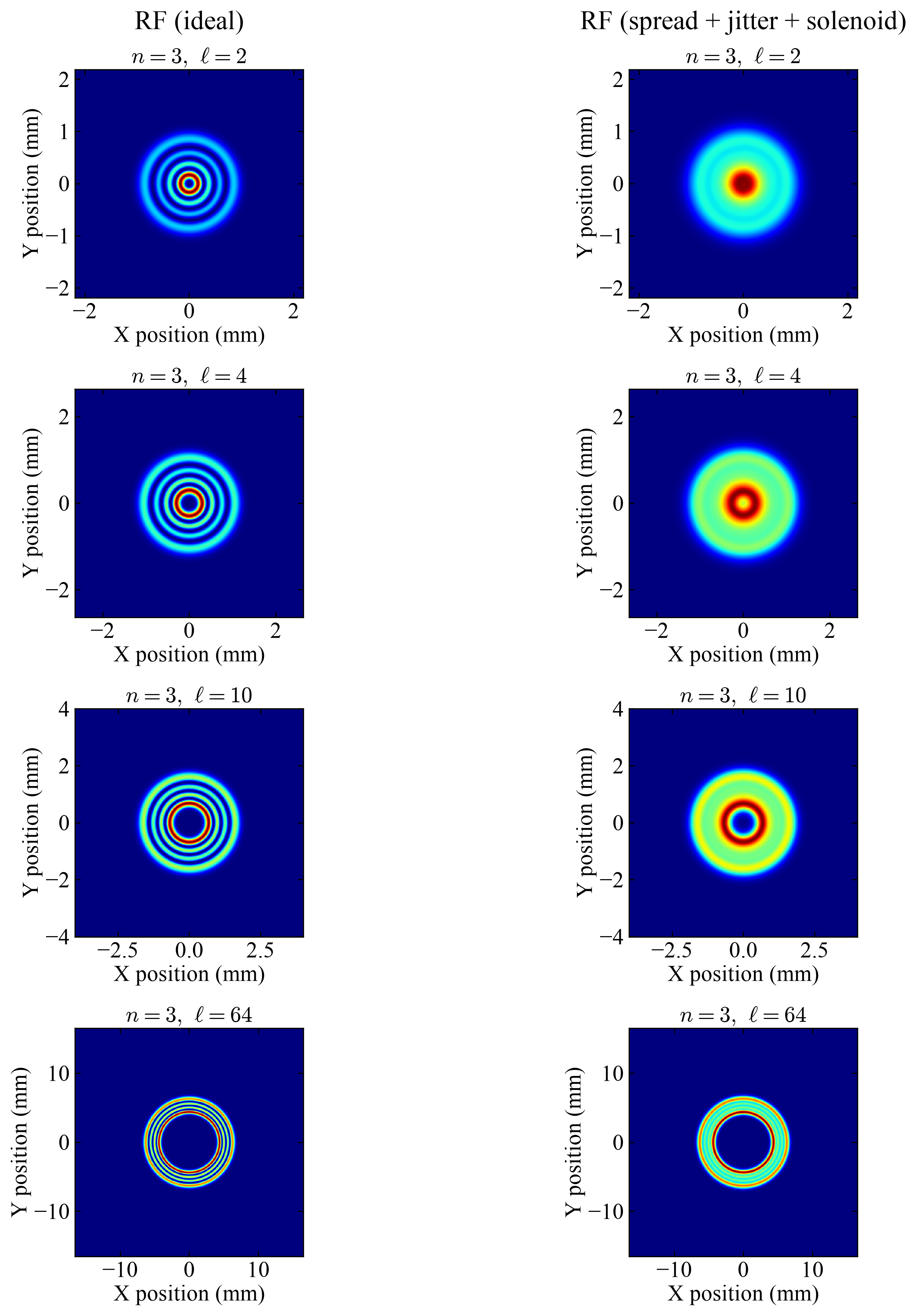}
        \caption{Dependence on the OAM value at the detector plane $z = 1.9$~m.}
        \label{fig:rf_oam_scan_nonideal}
    \end{subfigure}
    \hfill
    \begin{subfigure}[t]{0.45\textwidth}
        \centering
        \includegraphics[width=\textwidth]{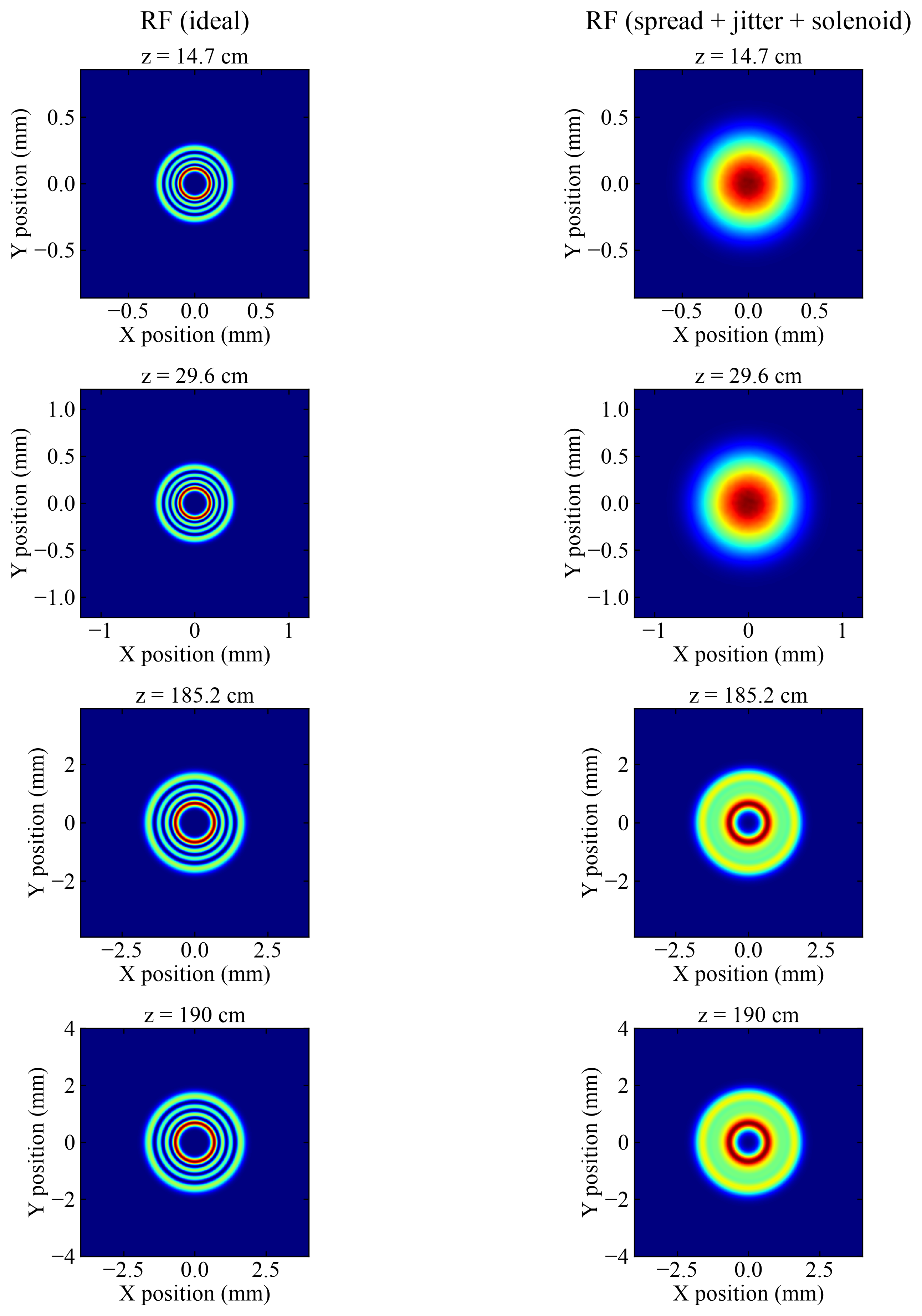}
        \caption{Dependence on the observation plane for the packet with $n=3$ and $\ell=10$.}
        \label{fig:rf_z_scan_nonideal}
    \end{subfigure}

    \caption{Transverse probability-density maps $|\psi(x,y)|^2$ in the ideal RF-acceleration model (left column) and the ASTRA-conditioned nonideal model (right column). Panel (a) shows the OAM dependence for 
    $n=3$ and $\ell=2,4,10,64$ at the detector plane $z=1.9$~m. Panel (b) shows the reference packet with $n=3$ and $\ell=10$ at $z=14.75$, $29.56$, $185.17$, and $190$~cm. The 
    nonideal maps include the ASTRA centroid distribution and RF amplitude and phase jitter. Each probability density is normalized to unit integral; the linear color scale is 
    adjusted independently for each panel.}
    \label{fig:rf_nominal_nonideal}
    \end{figure*}

    At $z=1.9$~m, the ensemble increases the total rms radius of the packet relative to the ideal RF case by approximately 4.5\%, 3.1\%, and 0.085\% for $\ell=2$, $10$,
    and $64$, respectively. The broadening is therefore most pronounced for the lower-OAM modes, while the large-$\ell$ annular structure remains comparatively stable
    (Fig.~\ref{fig:rf_nominal_nonideal}). These maps characterize intensity broadening; phase coherence and OAM conservation require phase-sensitive observables.
    
    For the present axisymmetric single-electron Hamiltonian, $[\hat{H}(t),\hat{L}_z]=0$; hence, for every RF amplitude and phase realization, a state with definite
    $\ell$ remains in the same OAM sector. Jitter changes its radial evolution and blurs the ensemble-averaged intensity. Non-axisymmetric field errors, collective interactions,
    and source decoherence can couple different OAM components and require a separate treatment.

    Photoemission statistics, finite emission time, and thermal cathode effects are fundamental sources of decoherence. Their quantitative description requires a
    many-body quantum-kinetic model of structured photoemission. The present single-electron calculation isolates transport in the prescribed RF field and predicts strong
    suppression of transverse spreading for an electron prepared in a definite $(n,\ell)$ state.

\section{\label{sec:conclusions}Conclusions}

    In this paper, we have presented the design of the JINR S-band RF photoinjector test bench, tailored specifically for the generation and acceleration of 
    relativistic vortex electron beams. To validate the proposed hardware configuration, we performed field-map-based classical and quantum beam-dynamics simulations. 
    Our tracking results demonstrate that the designed injector can operate stably in the low-charge regime ($Q = 0.63$~pC), where space-charge effects 
    remain subdominant after the initial acceleration stage. In this regime, the integrated compensation solenoid effectively controls 
    the transverse envelope and suppresses RF-driven phase-space distortions, yielding a reproducible operating point with a final normalized 
    transverse emittance of $56\ \pi \cdot \mathrm{\mu m \cdot mrad}$ that remains close to the intrinsic thermal-emittance scale of the photoemission source.

    Building upon this validated classical regime, we have studied the quantum evolution of Laguerre--Gaussian wave packets with realistic OAM values 
    up to $\ell = 64$ inside the designed accelerating fields. We showed that rapid energy gain in the realistic standing-wave RF field strongly suppresses transverse 
    quantum spreading by nearly three orders of magnitude (specifically, by a factor of $7.67 \times 10^3$ at the detector plane $z=1.9$~m) compared to free-space propagation. This 
    profound suppression ensures the preservation of the characteristic radial structure of the OAM-carrying states during acceleration.

    Overall, the presented design and simulations demonstrate that the current injector configuration fulfills the strict beam-quality 
    and ultra-high vacuum ($10^{-9}$--$10^{-8}$~Torr) requirements for experiments on the photoemission-driven generation of vortex electrons in the MeV range. We stress that this 
    process is extremely delicate: while macroscopic machine jitter and field imperfections do not fundamentally destroy the beam structure, fundamental sources of decoherence---such as 
    stochastic photoemission, thermal cathode effects, and residual multi-particle interactions---remain formidable challenges that highlight the complexity of preserving initial phase 
    coherence. The quantum calculations herein therefore serve as a single-electron baseline and an upper-bound estimate for the preservation of the transverse structure associated
    with the initial OAM state.

    In combination with a dedicated RF–laser synchronization system and tested diffractive optics for generating ultraviolet vortex beams up to $\ell = 64\hbar$, 
    the principal hardware components required for photoemission-based generation and transport of relativistic vortex electrons have been established and 
    characterized. Future studies will focus on the experimental commissioning of the test bench and the extension of the present theoretical modeling to include 
    collective decoherence mechanisms. The topological charge of the generated electron beams will be experimentally verified by diffraction on a triangular aperture, as has been 
    discussed in Ref.~\cite{Maksimov25}.

\section*{Acknowledgments}
    The authors express their gratitude to A.V.~Vukolov, M.V.~Shevelev, and G.K.~Sizykh for their significant assistance in developing the beam emittance diagnostics. We also 
    thank E.A.~Yakushev, V.V.~Glagolev, D. V.~Naumov, A.S.~Zhemchugov, and the management of the Laboratory of Nuclear Problems for supporting the project on behalf of JINR. 
    Special thanks are due to K.V.~Cherepanov and N.E.~Sheremet for their help with quantum dynamics calculations, and to S.S.~Baturin for valuable consultations. The authors 
    acknowledge A.V.~Afanasyev and N.I.~Garanzha for their valuable assistance in the assembly of the experimental setup, as well as M.A.~Shangareev for his help with the installation 
    and calibration of the laser beam transport line.
    
    This study was supported by the Russian Science Foundation (Project No. 23-62-10026)~\cite{RSF}.

\bibliographystyle{unsrtnat}
\bibliography{dynamics}

\end{document}